\documentclass[10pt,twocolumn,letterpaper]{article}

\usepackage[pagenumbers]{cvpr} 

\definecolor{cvprblue}{rgb}{0.21,0.49,0.74}
\usepackage[pagebackref,breaklinks,colorlinks,allcolors=cvprblue]{hyperref}
\usepackage{multirow}
\usepackage{braket}
\usepackage{subcaption}

\def\paperID{XXXXX} 
\def\confName{CVPR}
\def\confYear{2026}

\title{Quantum Implicit Neural Representations for 3D Scene Reconstruction and Novel View Synthesis}

\author{Y. Cordero\\
Computer Vision Center\\
Universitat Aut\`onoma de Barcelona\\
{\tt\small ycordero@cvc.uab.cat}
\and
P. García Molina\\
Computer Vision Center\\
Universitat Aut\`onoma de Barcelona\\
{\tt\small pgarciam@cvc.uab.cat}
\and
F. Vilariño\\
Computer Vision Center\\
Universitat Aut\`onoma de Barcelona\\
{\tt\small fernando@cvc.uab.es}
}

\begin{document}
\maketitle
\begin{abstract}
Implicit neural representations (INRs) have become a powerful paradigm for continuous signal modeling and 3D scene reconstruction, yet classical networks suffer from a well-known spectral bias that limits their ability to capture high-frequency details. Quantum Implicit Representation Networks (QIREN) mitigate this limitation by employing parameterized quantum circuits with inherent Fourier structures, enabling compact and expressive frequency modeling beyond classical MLPs. In this paper, we present Quantum Neural Radiance Fields (Q-NeRF), the first hybrid quantum–classical framework for neural radiance field rendering. Q-NeRF integrates QIREN modules into the Nerfacto backbone, preserving its efficient sampling, pose refinement, and volumetric rendering strategies while replacing selected density and radiance prediction components with quantum-enhanced counterparts. We systematically evaluate three hybrid configurations on standard multi-view indoor datasets, comparing them to classical baselines using PSNR, SSIM, and LPIPS metrics. Results show that hybrid quantum-classical models achieve competitive reconstruction quality under limited computational resources, with quantum modules particularly effective in representing fine-scale, view-dependent appearance. Although current implementations rely on quantum circuit simulators constrained to few-qubit regimes, the results highlight the potential of quantum encodings to alleviate spectral bias in implicit representations. Q-NeRF provides a foundational step toward scalable quantum-enabled 3D scene reconstruction and a baseline for future quantum neural rendering research.
\end{abstract}
    
\section{Introduction}
\label{sec:intro}

Novel view synthesis, the generation of realistic images of a scene from unseen viewpoints, is a central challenge in computer vision with applications in virtual reality~\cite{li2022immersiveneuralgraphicsprimitives}, autonomous navigation~\cite{wang2024nerfroboticssurvey}, and digital content creation~\cite{_lapak_2024}. Traditional 3D reconstruction methods based on explicit geometry such as meshes or point clouds \cite{schoenberger2016sfm} often fail to capture fine detail and view-dependent effects. Neural Radiance Fields (NeRF) \cite{mildenhall2020nerfrepresentingscenesneural} addressed these limitations by representing scenes as continuous volumetric functions that map 3D spatial locations and viewing directions to emitted color and density. By optimizing this representation through differentiable volume rendering, NeRF achieves photorealistic synthesis from sparse input views and has rapidly become a cornerstone in neural rendering. Subsequent variants improve scalability and reconstruction fidelity via multiscale representations \cite{barron2021mipnerfmultiscalerepresentationantialiasing, barron2022mipnerf360unboundedantialiased}, hash-grid encodings \cite{M_ller_2022}, and efficient sampling strategies \cite{reiser2021kilonerfspeedingneuralradiance}. Yet, despite their success, classical NeRFs remain computationally intensive and exhibit a pronounced \emph{spectral bias}, a tendency of MLP-based implicit networks to underfit high-frequency details \cite{tancik2020fourierfeaturesletnetworks}. This bias limits their ability to reproduce fine geometric structures and complex appearance variations that are essential for realistic rendering.

Quantum computing offers a complementary direction to overcome such representational bottlenecks. Quantum Neural Networks (QNNs) leverage superposition and entanglement to process information in exponentially large Hilbert spaces \cite{Biamonte_2017, Benedetti_2019}, potentially achieving greater expressivity per parameter than their classical counterparts. However, the training of QNNs is often hindered by the phenomenon of barren plateaus, regions in the optimization landscape where gradients vanish exponentially with system size, posing significant challenges to scalability and efficient learning \cite{McClean_2018, Cerezo_2021}. Parameterized quantum circuits (PQCs) naturally implement Fourier-like function decompositions \cite{schuld2021supervisedquantummachinelearning, Havl_ek_2019}, making them particularly suitable for modeling high-frequency components, precisely the regime where classical networks struggle. Recent studies in quantum implicit representations suggest that these circuits can compactly encode fine spectral details while maintaining trainability on noisy intermediate-scale quantum (NISQ) devices \cite{zhao2024quantumimplicitneuralrepresentations}. Such characteristics motivate exploring quantum-enhanced methods for continuous scene representations, bridging advances in quantum machine learning and neural rendering.

This work introduces Quantum Neural Radiance Fields (Q-NeRF), a hybrid quantum–classical framework that integrates Quantum Implicit Representation Networks (QIREN) \cite{zhao2024quantumimplicitneuralrepresentations} within the modular NeRF pipeline. Building on QIREN, Q-NeRF substitutes selected classical encoding or regression components with quantum modules while preserving the volumetric rendering formulation. This design enables controlled comparisons between classical and quantum components under identical rendering conditions. The framework is implemented in the NISQ regime through simulators that account for hardware constraints, shallow circuit depths, and hybrid optimization loops \cite{Endo_2021, Preskill2018quantumcomputingin}.

The main contributions of this paper are:
\begin{itemize}
    \item The Q-NeRF architecture, a modular hybrid NeRF architecture that incorporates QIREN-based quantum modules into a classical radiance field pipeline.
    \item A systematic comparison of three hybrid configurations, isolating the roles of quantum processing in density and color prediction.
    \item Empirical validation demonstrating the feasibility of integrating quantum circuits for high-frequency radiance modeling under current simulation constraints.
\end{itemize}

To our knowledge, Q-NeRF represents the first attempt to extend neural radiance fields into the quantum domain. This work establishes a foundation for studying how quantum computation can enhance implicit 3D representations, providing insights into the opportunities and limitations of hybrid quantum–classical models for future photorealistic scene reconstruction.

This work is structured as follows. \Cref{sec:related} reviews prior research on Neural Radiance Fields, Implicit Neural Representations, and Quantum Machine Learning, emphasizing advances in frequency modeling and hybrid quantum–classical training. \Cref{sec:method} details the proposed Q-NeRF architecture, describing its integration of QIREN modules into the NeRF pipeline and the three hybrid configurations explored. \Cref{sec:experiments} presents the experimental setup, baselines, and training procedures used to evaluate Q-NeRF under realistic simulation constraints. \Cref{sec:results} reports quantitative and qualitative results, analyzing reconstruction performance and the effects of quantum components. Finally, \Cref{sec:conclusions} concludes the paper and outlines future research directions.

\section{Related Work}
\label{sec:related}

Neural Radiance Fields (NeRF) formulate novel-view synthesis as learning a continuous 5D radiance field with volumetric rendering~\cite{mildenhall2020nerfrepresentingscenesneural}. Subsequent work improves antialiasing and robustness to scale with Mip-NeRF~\cite{barron2021mipnerfmultiscalerepresentationantialiasing} and Mip-NeRF-360~\cite{barron2022mipnerf360unboundedantialiased}, and accelerates training and inference via multi-resolution hash grids in Instant-NGP~\cite{M_ller_2022}. Alternative acceleration strategies include decomposing the scene into many small multilayer perceptron networks (MLPs) (KiloNeRF)~\cite{reiser2021kilonerfspeedingneuralradiance} and baking radiance into sparse/octree structures for real-time rendering (NSVF, PlenOctrees)~\cite{liu2021neuralsparsevoxelfields, yu2021plenoctreesrealtimerenderingneural}. These systems established the modular backbone we adopt (hash encodings, spherical harmonics for view direction, hierarchical sampling), upon which we integrate quantum modules. The technology can be applied in many areas, including real-world data capture~\cite{martinbrualla2021nerfwildneuralradiance}, robotics~\cite{wang2024nerfroboticssurvey}, and data compression~\cite{ pham2024neuralnerfcompression}.

Implicit Neural Representations (INRs) parameterize continuous signals with coordinate-based networks, with early roots in Compositional Pattern Producing Networks (CPPNs)~\cite{10.1007/s10710-007-9028-8} and universal approximation results for MLPs~\cite{Cybenko1989ApproximationBS}. Scene Representation Networks demonstrated learned implicit geometry and appearance~\cite{sitzmann2020scenerepresentationnetworkscontinuous}, but standard ReLU MLPs exhibit spectral bias toward low frequencies~\cite{rahaman2019spectralbiasneuralnetworks}. Random Fourier Features~\cite{NIPS2007_013a006f} and sinusoidal positional encodings~\cite{tancik2020fourierfeaturesletnetworks} mitigate this bias and underpin modern NeRF encoders.

Quantum Machine Learning (QML) explores quantum advantages for learning and generative modeling~\cite{Biamonte_2017, Carleo_2019}, typically implemented through variational quantum algorithms (VQAs) that operate in hybrid quantum–classical loops. In this pipeline, PQCs encode input data and generate quantum states whose measurement outcomes are evaluated by a classical optimizer, which iteratively updates the circuit parameters to minimize a loss function. This end-to-end training framework~\cite{Benedetti_2019, Schuld_2020} enables efficient integration of quantum computation for model evaluation with classical optimization for parameter tuning, forming the core of most practical QML architectures. Data re-uploading architectures alternate input encodings and trainable unitaries to increase expressivity~\cite{P_rez_Salinas_2020, Schuld_2021}, while quantum kernels and feature maps offer complementary approaches that project data into high-dimensional Hilbert spaces where complex, non-linear relationships can be more easily separated~\cite{Liu_2021, Huang_2021}. Practical challenges include barren plateaus and noise on NISQ devices~\cite{McClean_2018, Cerezo_2021, Bharti_2022}.

Quantum Implicit Neural Representations (QINRs) transfer INR principles to quantum settings by using PQCs as implicit function approximators. The QIREN realizes a Fourier-like decomposition through quantum encodings and expectation-value readout, enabling compact high-frequency representations~\cite{zhao2024quantumimplicitneuralrepresentations}. Extensions such as Quantum Fourier Gaussian Network (QFGN) further balance low-/high-frequency spectra with Fourier–Gaussian preprocessing before quantum encoding~\cite{jin2025qfgnquantumapproachhighfidelity}.
\section{Methodology}
\label{sec:method}

\begin{figure*}[t!]
    \centering
    \includegraphics[width=\textwidth]{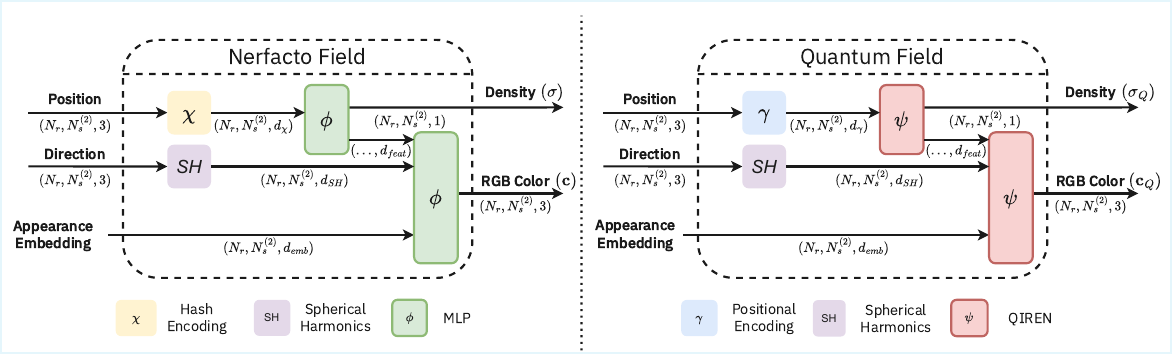}
    \caption{Comparison of rendering fields. The figure contrasts the classical Nerfacto field (left) with the Q-NeRF field (right). Both take spatial position $\mathbf{x}$, direction $\mathbf{d}$, and appearance embedding $\mathbf{a}$ as inputs, and produce density $\sigma$ and color $\mathbf{c}$.}
    \label{fig:nerfacto_field}
\end{figure*}

This section presents the proposed Q-NeRF architecture, a modular hybrid quantum–classical neural radiance field that extends Nerfacto \cite{Tancik_2023}, a state-of-the-art NeRF variant designed for efficient and modular 3D scene reconstruction. Q-NeRF preserves the core structure of the classical Nerfacto pipeline while replacing selected encoding or prediction components with quantum circuits, enabling direct comparison between hybrid configurations. The comparison between the classical Nerfacto field and our proposed Q-NeRF field is illustrated in Figure~\ref{fig:nerfacto_field}.

\subsection{Nerfacto Backbone}
\label{sec:nerfacto_backbone}
The baseline network consists of a fully connected multi-layer perceptron $F_{\theta}$ with shared parameters across all samples. The field separates into two prediction heads. The density branch, denoted \(F_{\theta}^{\text{dens}}\), processes only spatial information \(\mathbf{x}\) to prevent view leakage
\begin{equation}
\sigma = F_{\theta}^{\text{dens}}\!\bigl(\Phi(\mathbf{x})\bigr), \quad \sigma \ge 0,
\end{equation}
where \(\mathbf{x} = (x, y, z)\) denotes the 3D spatial coordinate, \(\Phi(\cdot)\) is the positional or hash encoding function, and \(\sigma \in \mathbb{R}_{\ge 0}\) represents the predicted volumetric density. The color branch conditions on spatial, directional, and appearance features
\begin{equation}
\mathbf{c} = F_{\theta}^{\text{rgb}}\!\bigl(\Phi(\mathbf{h}(\mathbf{x})), \mathrm{SH}(\mathbf{d}), \mathbf{a}\bigr)\in[0,1]^3.
\end{equation}
Here, \(\mathbf{d} \in \mathbb{R}^3\) is the normalized viewing direction, \(\mathbf{a}\) is the learned per-image appearance embedding, \(\mathrm{SH}(\cdot)\) denotes the spherical-harmonics encoder that captures angular variation, and $\mathbf{h}(\mathbf{x})$ denotes the intermediate geometric feature vector produced by the density branch (distinct from the scalar density $\sigma(\mathbf{x})$).

Both heads share initial layers, allowing geometry and appearance to co-regularise, while their last layers remain independent. ReLU activations are used in hidden layers, with softplus for $\sigma$ and sigmoid for $\mathbf{c}$ to ensure valid outputs.

To support unbounded scenes, input coordinates are contracted to the unit ball
\begin{equation}
\mathbf{x}_{\text{ctr}} =
\begin{cases}
\mathbf{x}, & \text{if } \|\mathbf{x}\|_2 \le 1,\\
\frac{2\mathbf{x}}{\|\mathbf{x}\|_2} - \frac{\mathbf{x}}{\|\mathbf{x}\|_2^2}, & \text{otherwise,}
\end{cases}
\end{equation}
where \(\|\mathbf{x}\|_2\) denotes the Euclidean norm, and the contraction ensures that unbounded world coordinates are mapped into a normalized unit sphere for stable hash-grid encoding.

The contracted coordinates $\mathbf{x}_{\text{ctr}}$ are fed to the encoding $\Phi$. The overall mapping
\begin{equation}
(\mathbf{x}, \mathbf{d}, \mathbf{a}) \longmapsto (\sigma, \mathbf{c})
\end{equation}
defines the Nerfacto field, which is integrated volumetrically for photorealistic rendering.

\subsection{Volumetric Rendering}
\label{sec:volumetric_rendering}
Given per-point predictions $(\sigma_i, \mathbf{c}_i)$ along a camera ray, the renderer computes transmittance-based weights
\begin{equation}
w_i = T_i(1 - e^{-\sigma_i \Delta t_i}),
\end{equation}
where $T_i = \prod_{j<i} \exp(-\sigma_j \Delta t_j)$ denotes the accumulated transmittance along the ray up to sample $i$, and $\Delta t_i$ is the distance between consecutive samples. The pixel color is then defined as
\begin{equation}
C(\mathbf{r}) = \sum_i w_i\, \mathbf{c}_i.
\end{equation}
This differentiable process allows gradient flow through the exponential attenuation and the softplus activation of $\sigma$. A hierarchical sampling strategy performs coarse-to-fine integration, refining intervals with high-frequency content.

\subsection{Quantum Field}
\label{sec:quantum_field}
The quantum field introduces parameterized quantum circuits (PQCs) into the radiance pipeline. Each PQC implements QIREN~\cite{zhao2024quantumimplicitneuralrepresentations}. QIREN replaces classical neural encodings by embedding input coordinates into higher-dimensional quantum feature spaces governed by a frequency parameter $\ell \in \{0,\dots,L\}$, which controls the highest represented frequency component. The architecture leverages the data re-uploading principle~\cite{P_rez_Salinas_2020}, enabling non-linear function approximation through alternating layers of data-encoding and trainable quantum unitaries, as demonstrated by using variational quantum circuits (VQCs)~\cite{Schuld_2021, Benedetti_2019}.  

To systematically study the role of quantum processing in radiance field modeling, we design three hybrid configurations that vary where QIREN is applied: only in color prediction, only in density prediction, or in both.

\subsubsection{Classical Density and Quantum Color}
\label{sec:hybrid_classicaldens_quantumcolor}
In the first configuration, QIREN handles color prediction, and density estimation remains classical.

The spatial position $\mathbf{x} \in \mathbb{R}^3$ is processed via a classical multi-resolution hash encoding for efficient spatial indexing
\begin{equation}
\sigma(\mathbf{x}) = \text{MLP}_{\text{classical}}\bigl(\gamma_{\text{hash}}(\mathcal{C}(\mathbf{x}))\bigr),
\end{equation}
where $\mathbf{d} \in \mathbb{R}^3$ and $\gamma_{\text{hash}}$ denote the viewing direction and classical hash-grid encoding \cite{M_ller_2022}, respectively, and $\mathcal{C}(\mathbf{x})$ denotes the standard scene-contraction mapping used in Nerfacto \cite{M_ller_2022}, 
which maps unbounded scene coordinates to the unit ball.
 
In contrast, the RGB color prediction employs QIREN to jointly encode position and viewing direction. Inputs $(\mathbf{x},\mathbf{d})$ are encoded into quantum states via data-dependent rotations applied to each qubit, forming an encoding unitary $U_{\text{enc}}(\mathbf{x},\mathbf{d})$.
The encoded data are processed by a PQC composed of parameterized single-qubit Euler rotations and entangling CZ gates
\begin{equation}
U_{\text{rot}}(\phi,\theta,\psi) = R_z(\psi)R_y(\theta)R_z(\phi),
\end{equation}
where $R_z$ and $R_y$ are single-qubit rotation gates about the $z$ and $y$ axes, respectively, 
and $(\phi, \theta, \psi)$ are trainable parameters of the circuit, as typically used in hardware-efficient quantum layers \cite{Benedetti_2019, Cerezo_2021}. The PQC follows a data re-uploading strategy \cite{P_rez_Salinas_2020}, where the encoded features $\gamma_{\text{Q}}(\mathbf{x}, \mathbf{d})$ are reintroduced at multiple layers to enrich the expressivity of the quantum model. 
We define the layered circuit with data re-uploading and entanglement as
\begin{equation}
U(\mathbf{x},\mathbf{d},\boldsymbol{\theta}) \;=\;
\prod_{l=1}^{L}\!\Big(\, U_{\mathrm{ent}}^{(l)} \; U_{\mathrm{rot}}(\boldsymbol{\theta}_l) \; U_{\mathrm{enc}}(\mathbf{h}(\mathbf{x}),\mathbf{d}) \,\Big),
\label{eq:layeredU}
\end{equation}
where $U_{\mathrm{enc}}(\mathbf{x},\mathbf{d})$ applies data-dependent single-qubit rotations encoding the input features
and $U_{\mathrm{ent}}^{(l)}$ is an entangling layer implemented as a ring of CZ gates, all parameters are represented as $\boldsymbol{\theta}$,
\(
U_{\mathrm{ent}}^{(l)}=\prod_{(i,j)\in\mathcal{E}} \mathrm{CZ}_{ij}
\). As in QIREN~\cite{zhao2024quantumimplicitneuralrepresentations}, the input $(\mathbf{x},\mathbf{d})$ is first processed by a classical linear layer to produce the rotation angles used in $U_{\mathrm{enc}}$, ensuring a differentiable hybrid pipeline. The resulting quantum embedding is obtained as the expectation value of a measurement observable $\mathcal{O}$,
\begin{equation}
\gamma_{\text{Q}}(\mathbf{x},\mathbf{d}) =
\langle 0| U^\dagger(\mathbf{x},\mathbf{d},\boldsymbol{\theta})
\mathcal{O}\,
U(\mathbf{x},\mathbf{d},\boldsymbol{\theta}) |0\rangle.
\label{eq:quantum_embedding}
\end{equation}
The expectation values of measurement observables from this circuit serve as the input features to the color prediction head $F_{\theta}^{\text{color}}$, such that
\begin{equation}
\mathbf{c}_Q(\mathbf{x},\mathbf{d}) = F_{\theta}^{\text{color}}\bigl(\gamma_{\text{Q}}(\mathbf{x},\mathbf{d})\bigr).
\label{eq:quantum_color}
\end{equation}

This hybrid setup integrates QIREN in the radiance branch, using its spectral expressivity for view-dependent effects while keeping the density path classical.

\subsubsection{Quantum Density and Classical Color}
\label{sec:hybrid_quantumdensity_classicalcolor}
In the second configuration, QIREN is used for density prediction, while color estimation remains classical.

Spatial coordinates $\mathbf{x}$ are encoded through PQCs with data re-uploading \cite{P_rez_Salinas_2020}, forming a QIREN density estimator
\begin{equation}
\sigma_Q(\mathbf{x}) =
\langle 0|
U^\dagger(\mathbf{x},\boldsymbol{\theta})
\mathcal{O}\,
U(\mathbf{x},\boldsymbol{\theta})
|0\rangle,
\label{eq:quantum_density}
\end{equation}
where $U(\mathbf{x},\boldsymbol{\theta}) = \prod_{l=1}^{L} [\,U_{\mathrm{ent}}^{(l)}\, U_{\text{rot}}(\boldsymbol{\theta}_l)\, U_{\text{enc}}(\mathbf{x})\,]$ 
represents the PQC composed of data-encoding unitaries $U_{\text{enc}}$, trainable layers $U_{\text{rot}}$ and entangling layers $U_{\mathrm{ent}}^{(l)}$, and $\mathcal{O}$ corresponds to a Pauli-$Z$ measurement on each qubit. The color is predicted using a classical MLP conditioned on geometric and directional features
\begin{equation}
\mathbf{c}(\mathbf{x},\mathbf{d}) = \text{MLP}_{\text{classical}}\bigl([\mathbf{h}(\mathbf{x}), \text{SH}(\mathbf{d})]\bigr),
\end{equation}
where $\text{SH}(\mathbf{d})$ encodes the viewing direction using spherical harmonics as in Nerfacto \cite{barron2021mipnerfmultiscalerepresentationantialiasing}.  

This variant isolates the contribution of quantum computation to spatial density estimation, emphasizing geometry learning while preserving a lightweight, interpretable classical appearance model.

\subsubsection{Quantum Density and Quantum Color}
\label{sec:hybrid_quantum_all}
In the third configuration, QIREN modules are employed in both the density and color branches, integrating quantum processing throughout the entire radiance field pipeline.

The quantum embeddings and the corresponding readout mappings for density and color
are defined in Eqs.~\eqref{eq:quantum_embedding}, \eqref{eq:quantum_color}, and
\eqref{eq:quantum_density}. The overall radiance field output is then given by
\begin{equation}
F_{\theta}(\mathbf{x},\mathbf{d}) = \sigma_Q(\mathbf{x}) + \mathbf{c_Q}(\mathbf{x},\mathbf{d}),
\label{eq:quantum_field_output}
\end{equation}
where $\sigma_Q(\mathbf{x})$ and $\mathbf{c_Q}(\mathbf{x},\mathbf{d})$ denote the
quantum-predicted density and color components, respectively.

This configuration integrates quantum computation into both geometry and radiance estimation, enabling a comprehensive analysis of quantum contributions to scene representation. 
Although classical simulation of quantum layers remains computationally expensive, this design serves as a conceptual benchmark for hybrid integration, highlighting the computational bottlenecks of simulating quantum circuits on classical hardware \cite{Preskill2018quantumcomputingin, Endo_2021}.

\begin{table*}[t]
\centering
\begin{tabular}{llcccccc}
\toprule
\textbf{Type} & \textbf{Config} & \textbf{\# Layers} & \textbf{Layer Width} & \textbf{Params} & \textbf{PSNR $\uparrow$} & \textbf{SSIM $\uparrow$} & \textbf{LPIPS $\downarrow$} \\
\midrule
\multirow{9}{*}{\textbf{Quantum (Q-NeRF)}} 
 & QIREN & 3L+4S & 8  & 1,339 & 30.49 & 0.93 & 0.0087 \\
 & QIREN & 3L+2S & 8  &   955 & 29.31 & 0.92 & 0.0171 \\
 & QIREN & 3L+1S & 8  &   763 & 29.33 & 0.93 & 0.0078 \\
 & QIREN & 2L+4S & 8  & 1,011 & 29.63 & 0.93 & 0.0107 \\
 & QIREN & 2L+2S & 8  &   723 & 28.54 & 0.92 & 0.0106 \\
 & QIREN & 2L+1S & 8  &   579 & 29.89 & \textbf{0.94} & 0.0073 \\
 & QIREN & 1L+4S & 8  &   683 & 28.64 & 0.92 & 0.0094 \\
 & QIREN & 1L+2S & 8  &   491 & \textbf{30.74} & \textbf{0.94} & \textbf{0.0067} \\
 & QIREN & 1L+1S & 8  &   395 & 30.05 & \textbf{0.94} & 0.0073 \\
\midrule
\multirow{14}{*}{\textbf{Classical (Nerfacto)}} 
 & MLP & 8  & 10 &   913   & 14.29 & 0.23 & 0.2720 \\
 & MLP & 8  & 12 &  1,239  & 14.52 & 0.27 & 0.2300 \\
 & MLP & 8  & 16 &  2,035  & 15.48 & 0.27 & 0.2470 \\
 & MLP & 8  & 32 &  7,139  & 14.97 & 0.26 & 0.2070 \\
 & MLP & 3  & 10 &    363  & 30.83 & \textbf{0.95} & 0.0061 \\
 & MLP & 3  & 12 &    459  & 30.86 & \textbf{0.95} & 0.0058 \\
 & MLP & 3  & 16 &    675  & 30.69 & 0.94 & 0.0056 \\
 & MLP & 3  & 32 &  5,763  & \textbf{31.12} & \textbf{0.95} & \textbf{0.0051} \\
 & MLP & 3  & 64 & 26,563  & 15.19 & 0.27 & 0.1600 \\
 & MLP & 4  & 12 &    771  & 30.07 & 0.94 & 0.0066 \\
 & MLP & 5  & 12 &    927  & 30.64 & \textbf{0.95} & 0.0056 \\
 & MLP & 6  & 12 &  1,083  & 16.34 & 0.36 & 0.1740 \\
 & MLP & 7  & 12 &  1,239  & 16.36 & 0.38 & 0.2190 \\
 & MLP & 8  & 12 &  1,239  & 14.52 & 0.27 & 0.2300 \\
\bottomrule
\end{tabular}
\caption{Performance comparison between Quantum (QIREN) and Classical (MLP) RGB predictors in the Q-NeRF and Nerfacto architectures. Each configuration specifies the number of layers, layer width, and total parameters in the color network.}
\label{tab:full_results}
\end{table*}

\section{Experimental Setup}
\label{sec:experiments}

\subsection{Implementation \& Dataset}
The evaluation of the Q-NeRF framework focuses on assessing the feasibility of integrating quantum components into a NeRF pipeline under practical simulation and computational constraints.

All experiments are performed on a controlled indoor scene to isolate the effects of the hybrid architecture from environmental variability and data diversity. The experiments employ the poster scene from the Nerfstudio dataset collection \cite{Tancik_2023}, which contains multi-view RGB captures with consistent illumination and calibrated camera poses. The scene includes planar structures and reflective textures that challenge both geometric and photometric modeling. Images are downsampled to 36 × 64 pixels to maintain spatial consistency while reducing the computational load required for quantum simulation. A 90\%-10\% split is applied for training and evaluation, ensuring that test views remain unseen during optimization. This configuration preserves the full range of viewing angles while keeping training time manageable.

All models are implemented within the Nerfstudio framework using PyTorch for classical components and PennyLane for quantum circuit simulation. The Adam optimizer is employed with an initial learning rate of $10^{-2}$, decayed exponentially to $10^{-4}$ over 30,000 iterations. Each training run maintains identical scheduling, sampling, and regularization parameters across all configurations to ensure methodological consistency. Two proposal-sampling stages are used per ray (256 and 96 samples), with proposal-weight annealing during the first 1,000 iterations.

To accommodate the computational limits of quantum simulation, the number of rays per batch is reduced to 128. Quantum modules are simulated using a state-vector backend with up to eight qubits and five data re-uploading layers, representing the maximum configuration feasible on standard GPU hardware while retaining convergence stability.


\subsection{Baselines and Hybrid Configurations}
The study compares the proposed Q-NeRF variants against the classical Nerfacto implementation, following an identical optimization pipeline and loss formulation. All models share the same sampling strategy, number of iterations, and dataset split to ensure experimental consistency.

The classical Nerfacto model serves as the primary reference. It combines multi-resolution hash encoding with a fully connected multilayer perceptron to predict both volumetric density and emitted color from spatial coordinates and view directions. This design provides strong baseline performance and is representative of recent implicit radiance field architectures used in high-quality scene reconstruction. For fair comparison with the quantum density variants of Q-NeRF, the baseline instead employs positional encoding rather than hash encoding, ensuring that performance differences arise from the quantum modules rather than the input representation. An ablation comparing classical Nerfacto variants with and without hash encoding is included in the supplementary material.




\subsection{Traning Procedure and Stability}

Each training iteration samples a batch of rays from the training images and estimates pixel colors through volumetric rendering. The loss function combines an $L_2$ photometric error with regularization on transmittance weights to stabilize early-stage convergence. Quantum circuits are differentiated using the parameter-shift rule \cite{Mitarai_2018, Schuld_2019_paramshift}, which allows backpropagation through the quantum layer while preserving numerical stability. Gradients from quantum modules are seamlessly integrated into the automatic differentiation graph of the classical NeRF pipeline, ensuring unified optimization across both domains.


Each iteration requires approximately 2.2 seconds on a single NVIDIA RTX 3090 GPU, resulting in a total training time of roughly 18 hours for the hybrid configurations. The classical baseline completes in approximately 5 hours under identical conditions. On future fault-tolerant quantum hardware, PQCs could exploit quantum parallelism and high-dimensional state evolution to evaluate forward passes and gradients more efficiently than their classical counterparts \cite{Biamonte_2017, Cerezo_2021}. Thus, while simulation increases training time, the same architecture should gain significant speedups on real quantum devices.


\section{Results \& Discussion}
\label{sec:results}

We evaluate the proposed Q-NeRF architecture using PSNR~\cite{Netravali1988DigitalPictures}, SSIM~\cite{wang2004image}, and LPIPS~\cite{zhang2018unreasonable}. Results are averaged over held-out test views to assess numerical fidelity, structural similarity, and perceptual realism.

\subsection{Classical Density + Quantum Color}
We evaluate Q-NeRF variants where the QIREN module replaces only the RGB predictor, as defined in \cref{sec:hybrid_classicaldens_quantumcolor}. The complete results for these RGB-only variants are shown in~Table~\ref{tab:full_results}. Here, the notation “$x$L+$y$S” refers to architectures with $x$ QIREN layers and $y$ spectral (data re-uploading) layers per QIREN layer. Compact Q-NeRF predictors, such as 1L+2S (491 params) and 1L+1S (395 params), achieve 30.74 dB and 30.05 dB PSNR with SSIM=0.94 (Table~\ref{tab:full_results}), demonstrating stable and high-quality reconstructions despite small model size.

In contrast, classical Nerfacto predictors show high variance in performance. Some lightweight models reach 30 dB PSNR and SSIM around 0.95, while others of similar size collapse below 15 dB with poor structural similarity. This variance shows that scaling classical Nerfacto does not reliably improve quality: some settings (e.g., 3 layers, width 10–32) reach $\sim$31 dB with SSIM $\approx$ 0.95, whereas nearby configurations (e.g., 7–8 layers with small widths) collapse to $\sim$14–16 dB and poor SSIM, indicating a less robust parameter–performance relationship than Q-NeRF.

At comparable small parameter budgets ($\sim$400–1,000 params), Q-NeRF delivers stable $\sim$29–31 dB performance, whereas Nerfacto range from $\sim$14 dB to $\sim$31 dB. For example, Q-NeRF 3L+2S (955 params) reaches 29.31 dB, while a Nerfacto of similar size (913 params) collapses to 14.29 dB; conversely, other small Nerfacto configurations (e.g., 3 layers, 363–459 params) attain $\sim$30.8–30.9 dB. Thus, Q-NeRF does not uniformly surpass the best Nerfacto in PSNR, but it avoids the frequent performance collapses observed in nearby Nerfacto configurations, providing more stable behavior across model scales. The sharp degradation in the Nerfacto arises from optimization instability and sensitivity to initialization, reflecting the weaker inductive bias of classical networks under tight parameter budgets.

\begin{figure*}[t!]
    \centering

    \begin{subfigure}[t]{0.49\linewidth}
        \centering
        \includegraphics[width=\linewidth]{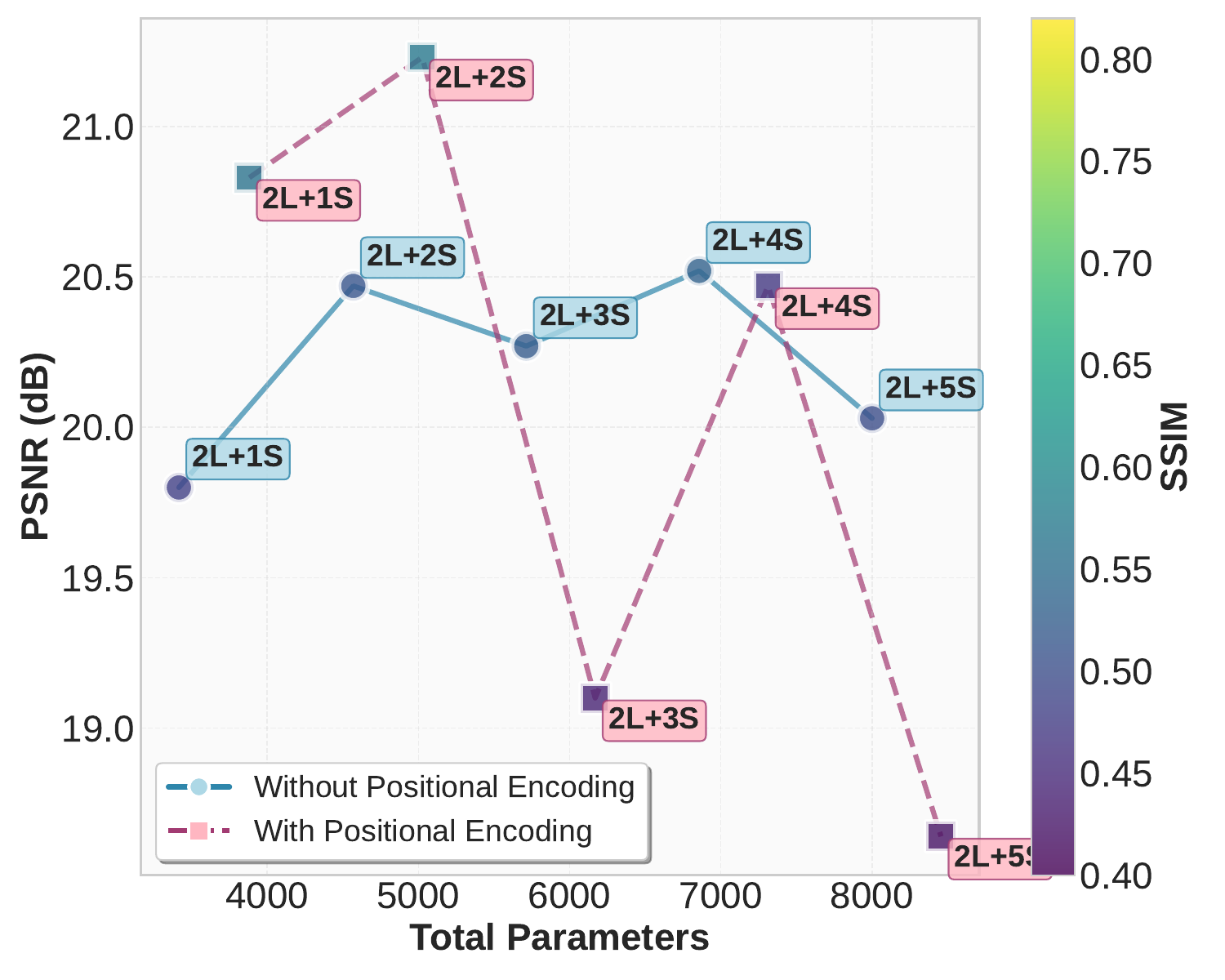}
        \caption{Quantum density Q-NeRF variant. PSNR vs.\ total parameter count; QIREN layers are denoted “L” and spectral (data re-upload) layers “S.”}
        \label{fig:quantum_density}
    \end{subfigure}\hfill
    \begin{subfigure}[t]{0.49\linewidth}
        \centering
        \includegraphics[width=\linewidth]{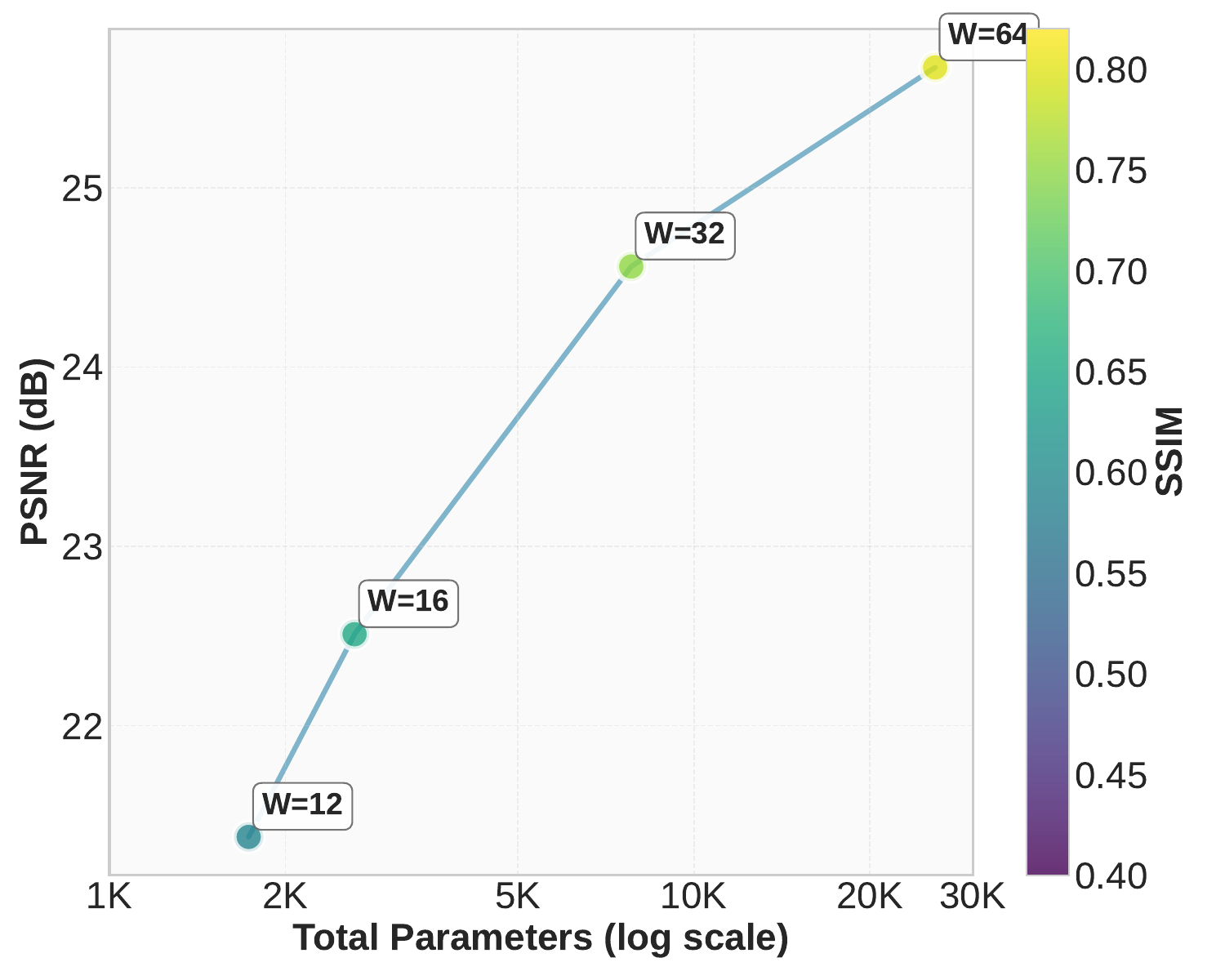}
        \caption{Classical Nerfacto with positional encoding. Average validation PSNR for varying hidden widths.}
        \label{fig:pos_encoding_analysis}
    \end{subfigure}

    \caption{Architecture–capacity comparison: both panels plot PSNR, with marker color indicating SSIM.}
    \label{fig:quantum_vs_classical}
\end{figure*}

\begin{table}[t]
\centering
\small
\begin{tabular}{llccccc}
\toprule
\textbf{D} & \textbf{RGB} & \textbf{Params} & \textbf{PSNR} & \textbf{SSIM} & \textbf{LPIPS} \\
\midrule
2L+1S & 1L+1S & 566/395 & 19.53 & 0.48 & 0.445 \\
2L+2S & 1L+2S & 710/491 & 19.70 & 0.49 & 0.404 \\
2L+3S & 1L+3S & 854/587 & 19.92 & 0.50 & 0.376 \\
2L+4S & 1L+4S & 998/683 & \textbf{20.24} & \textbf{0.51} & \textbf{0.329} \\
2L+5S & 1L+5S & 1142/779 & 19.01 & 0.45 & 0.490 \\
2L+6S & 1L+6S & 1286/875 & 19.28 & 0.34 & 0.620 \\
3L+1S & 1L+1S & 566/395 & 19.53 & 0.47 & 0.410 \\
4L+1S & 3L+1S & 750/579 & 19.60 & 0.48 & 0.347 \\
5L+1S & 4L+1S & 934/763 & 18.82 & 0.44 & 0.526 \\
4L+4S & 3L+4S & 1118/1339 & 16.77 & 0.36 & 0.534 \\
\bottomrule
\end{tabular}
\caption{Performance of the quantum density and quantum RGB Q-NeRF variant using different QIREN configurations. Each model includes both density (D) and color (RGB) QIREN modules. Params shows density/RGB parameter counts.}
\label{tab:qiren_results}
\end{table}

\subsection{Quantum Density + Classical Color}
In this configuration, QIREN replaces the density branch while the RGB color head remains fully classical using spherical harmonics, as in \cref{sec:hybrid_quantumdensity_classicalcolor}. The corresponding results, shown in Fig.~\ref{fig:quantum_density}, indicate that compact quantum density modules substantially improve geometric representation up to a certain model size. The best-performing configuration, 2L+2S, achieves 21.23 dB PSNR when combined with positional encoding, outperforming wider or deeper variants. However, performance degrades for larger models such as 2L+5S, suggesting diminishing returns and potential overfitting as circuit depth increases.

When compared to the classical Nerfacto baseline~(Fig.~\ref{fig:pos_encoding_analysis}), which achieves around 25.67~dB PSNR for wide MLPs (W=64) with positional encoding, the quantum density branch currently underperforms in absolute reconstruction accuracy. 
Under comparable parameter budgets, the Q-NeRF density branch attains PSNR/SSIM that are broadly similar to the classical positional-encoding baseline~(Fig.~\ref{fig:quantum_vs_classical}), though it does not surpass it in absolute accuracy. These results suggest that compact Q-NeRF modules can reach comparable density performance to classical models of similar size, though without providing a clear advantage yet.

\begin{figure}[t!]
    \centering
    \includegraphics[width=\linewidth]{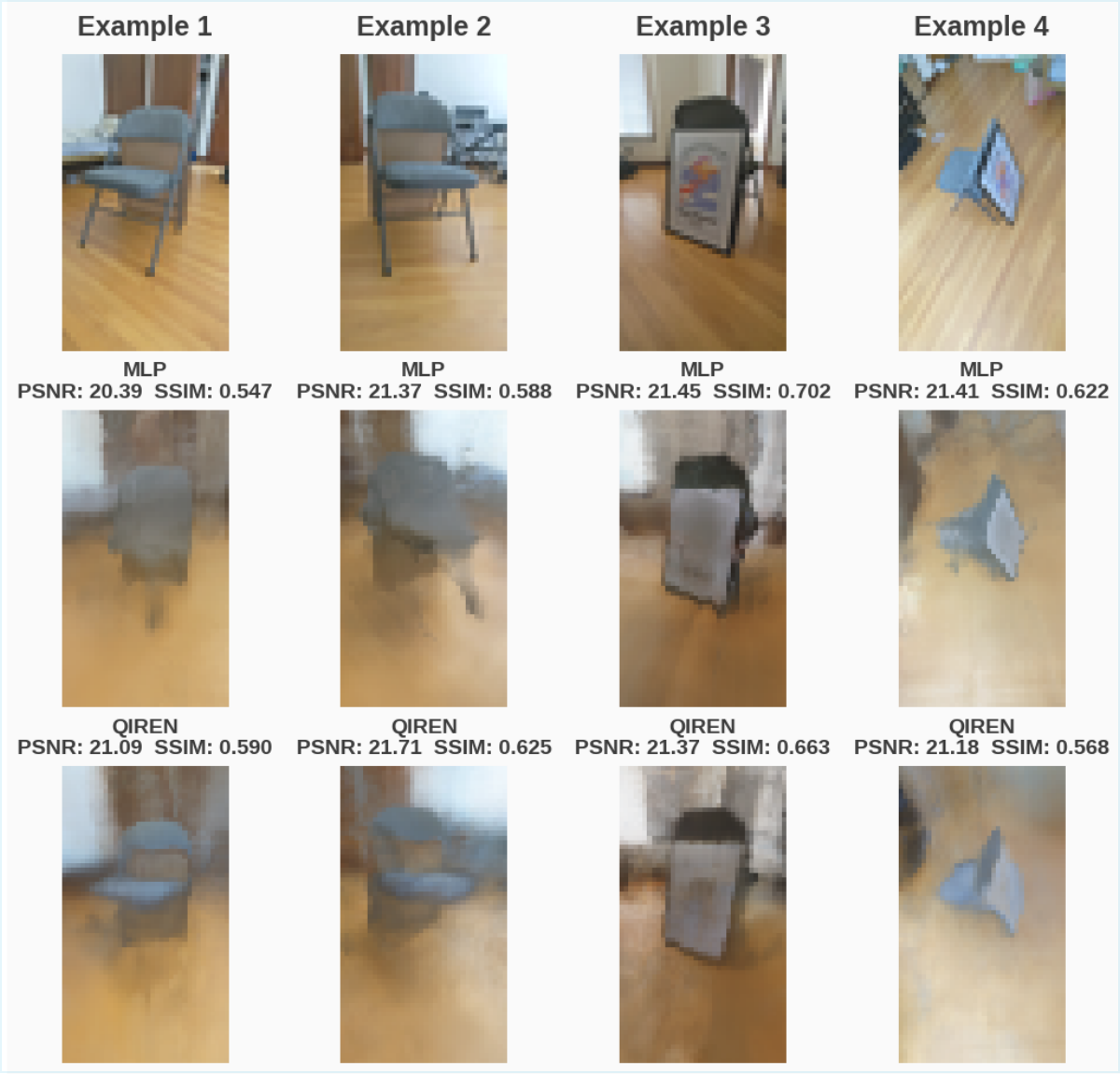}
    \caption{Qualitative comparison between the best-performing quantum-based density predictor (QIREN) and a comparable classical MLP-based density predictor. Each column shows a test-set example, with the top row depicting MLP predictions and the bottom row showing QIREN predictions.}
    \label{fig:qualitative_density}
\end{figure}

\subsection{Quantum Density + Quantum Color}
Table~\ref{tab:qiren_results} summarizes the performance of Q-NeRF variants where both the density and RGB components use QIREN modules (\cref{sec:hybrid_quantum_all}). 
Increasing the number of spectral layers in the density branch improves reconstruction quality up to a moderate configuration, with ``2L+4S (D) + 1L+4S (RGB)'' achieving the best scores (20.24~dB PSNR, 0.51 SSIM, 0.329 LPIPS). 
Beyond this point, deeper configurations such as ``2L+5S'' and ``2L+6S'' do not yield further gains, suggesting that larger quantum circuits do not necessarily translate to higher-quality reconstructions under the same training setup.

Compared with the classical Nerfacto baseline using positional encoding (Fig.~\ref{fig:pos_encoding_analysis}), which reaches 25.67~dB PSNR with approximately 30k parameters, the Q-NeRF variant attains lower absolute accuracy but does so with nearly an order of magnitude fewer parameters ($\approx$ 1k).
This highlights that while the quantum model is less accurate in absolute terms, it achieves comparable qualitative reconstructions with substantially reduced model complexity, illustrating the efficiency potential of hybrid quantum architectures.

\subsection{Qualitative Comparison}



Figure~\ref{fig:qualitative_density} presents a qualitative comparison between the best-performing quantum density variant of Q-NeRF and a size-matched Nerfacto on unseen test views. Both models exhibit similar global appearance, but Q-NeRF tends to preserve sharper edges and fine-grained textures, whereas Nerfacto outputs appear smoother and slightly blur detailed regions. In some cases (e.g., Scenes 1–2), Q-NeRF attains higher SSIM, showing better structural consistency, while in others (Scenes 3–4) Nerfacto yields slightly higher PSNR from its smoother reconstruction.

Visual inspection confirms that Q-NeRF occasionally introduces mild high-frequency artifacts but consistently maintains sharper local detail, while the Nerfacto exhibits homogeneous blur that suppresses fine geometry. This trade-off suggests that Q-NeRF’s spectral flexibility aids detail reconstruction even when pixel-wise error metrics slightly favor smoother outputs. Overall, the quantum-informed predictor achieves perceptual quality competitive with, and in certain cases superior to, its classical counterpart despite using a comparable number of parameters and operating under identical volumetric rendering conditions.

\subsection{Discussion}
The results indicate that Q-NeRF can reproduce meaningful radiance field reconstructions even with compact quantum modules simulated on classical hardware. Within the tested range (hundreds to roughly one thousand parameters), Q-NeRF variants achieve reconstruction quality comparable to classical counterparts, suggesting that quantum feature encodings can serve as efficient alternatives for low-capacity models. Although the explored configurations remain much smaller than typical classical baselines, the experiments demonstrate the feasibility of integrating PQCs into NeRF pipelines and provide a foundation for future work on larger and higher-quality hybrid architectures. These findings suggest that quantum encodings can complement classical NeRFs by introducing spectral diversity and implicit regularization, even under limited simulation budgets.

Among the tested configurations, the Q NeRF variant using a quantum color predictor and a classical density branch achieved the most competitive results. This behavior stems from combining the classical hash encoding, which captures geometric structure efficiently, with the quantum color module, which enhances view-dependent appearance.

The experimental setup serves as a controlled proof of concept for integrating quantum components into a NeRF pipeline, with all quantum operations simulated classically for precise benchmarking and stable gradients. Although this setup is computationally constrained, it offers a reliable environment to validate the feasibility and stability of hybrid quantum–classical architectures.

While real quantum processors were not used in this study, the simulated framework mirrors algorithms compatible with NISQ devices. As hardware improves in coherence, fidelity, and connectivity, Q-NeRF’s design may provide intrinsic advantages over classical methods. In particular, PQCs can represent high-frequency components with exponentially fewer parameters \cite{schuld2021supervisedquantummachinelearning, Havl_ek_2019}, and their native access to large Hilbert spaces may enable more compact and expressive radiance representations.


\section{Conclusions \& Future Work}
\label{sec:conclusions}

This work introduced Q-NeRF, a hybrid quantum and classical framework that incorporates PQCs into the NeRF rendering pipeline to investigate quantum enhanced 3D scene reconstruction. Embedding QIREN within the Nerfacto architecture enabled the modeling of fine scale, view dependent features while maintaining competitive reconstruction quality under strict computational constraints. Experiments on simulated quantum environments, limited to small qubit counts, provided one of the first systematic evaluations of quantum components in NeRFs.

The results indicate that hybrid quantum models can preserve visual fidelity with substantially fewer parameters, suggesting a potential route toward more compact and expressive implicit representations. Nonetheless, the reliance on classical simulation currently imposes clear limitations on scalability, training stability, and runtime efficiency. In a post-NISQ era, with access to fault-tolerant quantum processors and larger qubit counts, these architectures could be executed natively on quantum hardware, enabling direct benchmarking of their representational advantages without classical simulation overhead.

Future research should focus on deploying Q-NeRF on near-term quantum hardware, exploring error-mitigated and variational circuit designs tailored to radiance field learning, and improving hybrid training efficiency through parameter sharing and batching strategies. Extending quantum integration to dynamic scenes, Gaussian splatting, and other implicit representation frameworks constitutes a promising step toward scalable quantum-based neural rendering. Furthermore, future evaluations should include more complex and realistic datasets to better assess the model's generalization in practical scenarios.

\section*{Acknowledgments}
This piece of research was carried out with the support of the following grant projects: XXXXXX, XXXXXX, and XXXXXX [REMOVED FOR PEER REVIEW].

\section*{Code Availability}
The code will be made publicly available after peer review.



{
    \small
    \bibliographystyle{ieeenat_fullname}
    \bibliography{main}
}

\clearpage
\setcounter{page}{1}
\maketitlesupplementary

\appendix

\section{Introduction to Quantum Computing}
\label{sec:appendix_intro_quantum}

This section introduces fundamental quantum computing concepts integral to QML, specifically targeted at the computer science audience, with mathematical formulations and common applications to bridge understanding between classical and quantum paradigms.

Dirac notation, employing the symbol $\ket{\psi}$ known as a "ket", provides a convenient way to denote quantum states without referring to the particular function used to represent them~\cite{Nielsen_Chuang_2010}. This mathematical formalism, developed by Paul Dirac, forms the foundation for expressing quantum mechanical operations and is essential for understanding quantum machine learning algorithms. The notation consists of two fundamental components: kets and bras. Kets represent quantum states $\ket{\psi}$ which correspond to vectors in a complex Hilbert space $\mathcal{H}$. Mathematically, a ket can be expressed as
\[
\ket{\psi} = \sum_i c_i \ket{i},
\]
where $\{c_i\}$ are complex coefficients and $\{\ket{i}\}$ form an orthonormal basis of the Hilbert space. Bras are defined as the complex conjugate transpose of kets. For a ket $\ket{\psi} = \sum_i c_i \ket{i}$, the corresponding bra is
\[
\bra{\psi} = \sum_i c_i^* \bra{i},
\]
where $c_i^*$ denotes the complex conjugate of $c_i$. The bra-ket notation also defines scalar products between quantum states. The inner product between two states $\ket{\phi} = \sum_i d_i \ket{i}$ and $\ket{\psi} = \sum_j c_j \ket{j}$ is given by
\[
\braket{\phi|\psi} = \sum_{i,j} d_i^* c_j \braket{i|j} = \sum_i d_i^* c_i,
\]
where the orthonormality condition $\braket{i|j} = \delta_{ij}$ has been used. Finally, the notation includes outer products, which represent operators. The outer product of two states $\ket{\psi}$ and $\bra{\phi}$ is written as
\[
\ket{\psi}\bra{\phi} = \sum_{i,j} c_i d_j^* \ket{i}\bra{j}.
\]
These operators are widely used in quantum mechanics and quantum computing for constructing density matrices and projection operators.

\begin{figure}[t]
    \centering
    \includegraphics[width=0.3\textwidth]{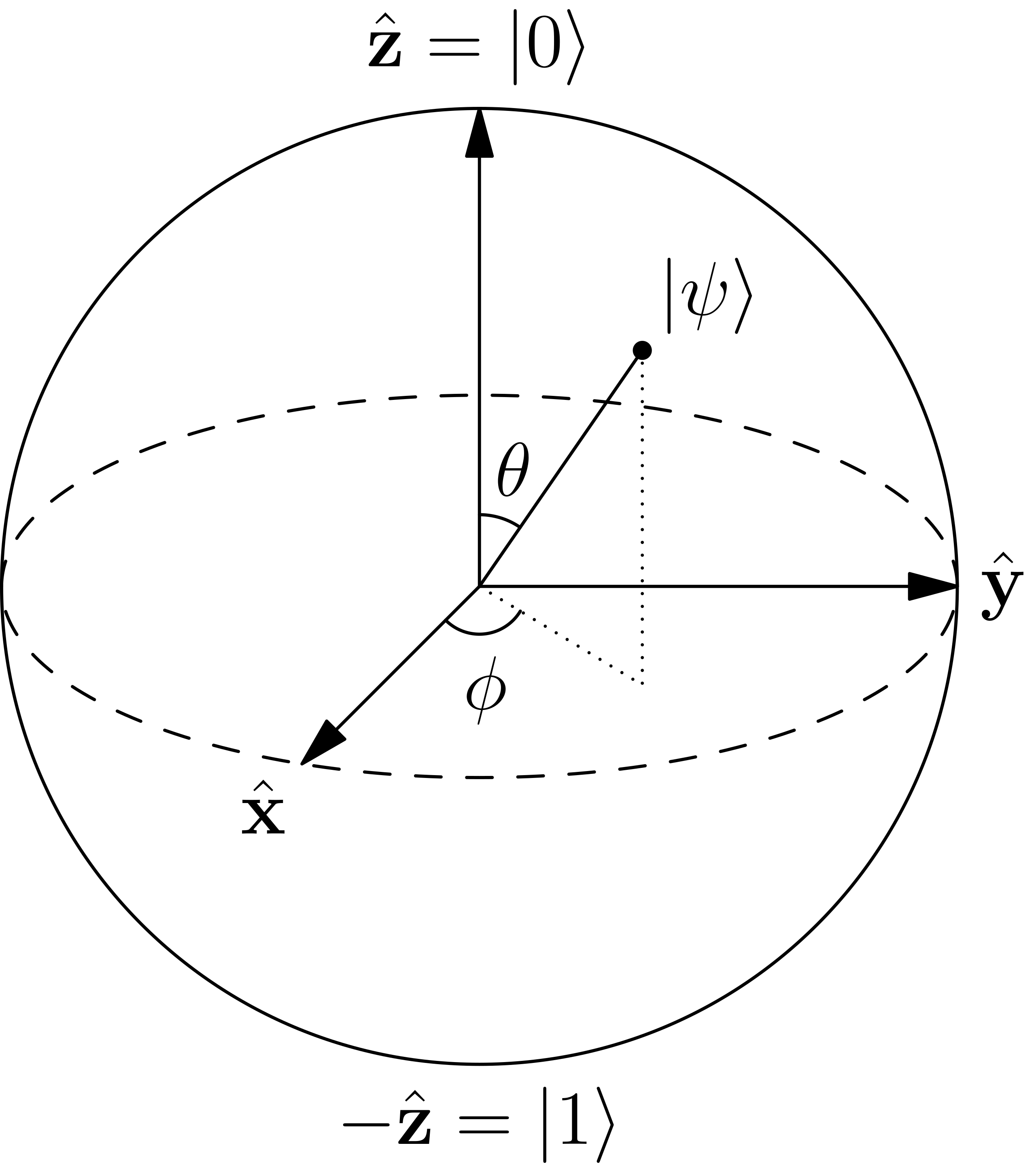}
    \caption{The Bloch sphere representation of a qubit state. The angles $\theta$ and $\phi$ define the position of the state vector on the sphere, encoding the superposition and relative phase of the basis states $\ket{0}$ and $\ket{1}$.}
    \label{fig:bloch_sphere}
\end{figure}

A qubit, or quantum bit, is the fundamental unit of quantum information. Unlike a classical bit, which is represented as either $ 0 $ or $ 1 $, a qubit exists in a superposition of the basis states $ \ket{0} $ and $ \ket{1} $:
$
\ket{\psi} = \alpha \ket{0} + \beta \ket{1}, \quad \text{where} \quad |\alpha|^2 + |\beta|^2 = 1.
$
Here, $ \alpha, \beta \in \mathbb{C} $ are complex amplitudes that determine the probabilities of measuring $ \ket{0} $ and $ \ket{1} $, respectively. This probabilistic nature lies at the heart of quantum computation \cite{Nielsen_Chuang_2010}.

Quantum measurement collapses a quantum state $ \ket{\psi} $ into one of its basis states. The probability of measuring a state $ \ket{i} $ is given by
\begin{equation}
P(\ket{i}) = |\braket{i|\psi}|^2.
\end{equation}
Measurement outcomes are fundamental to extracting classical information from a quantum system and are central to the functioning of hybrid quantum-classical models.

Superposition allows a single qubit to represent multiple states simultaneously. For example, a superposition state $ \frac{1}{\sqrt{2}}(\ket{0} + \ket{1}) $ represents an equal probability of measuring either $ \ket{0} $ or $ \ket{1} $, where each outcome has probability $|\frac{1}{\sqrt{2}}|^2 = \frac{1}{2}$. In systems with multiple qubits, entanglement emerges as a key quantum phenomenon, where the quantum state of one qubit is dependent on another, even if spatially separated
\begin{equation}
\ket{\psi_{\text{entangled}}} = \frac{1}{\sqrt{2}}(\ket{00} + \ket{11}).
\end{equation}
Such correlations cannot be explained by classical physics and are pivotal in QML algorithms and methods.

The Bloch sphere provides a geometric representation of a qubit as a point on a unit sphere in 3D space. Any qubit state $ \ket{\psi} = \cos(\theta/2) \ket{0} + e^{i\phi} \sin(\theta/2) \ket{1} $ can be represented on this radius-1 sphere through angles $ \theta $ (polar) and $ \phi $ (azimuthal). This visualization succinctly captures superposition and phase relationships and is widely used to illustrate quantum state evolution under quantum gates (see Figure~\ref{fig:bloch_sphere}).

Quantum gates are unitary operations that manipulate qubits, analogous to logical gates in classical computation. Common single-qubit gates include
\[
\begin{aligned}
X &= 
\begin{bmatrix}
0 & 1 \\
1 & 0
\end{bmatrix}, \quad
H = \frac{1}{\sqrt{2}}
\begin{bmatrix}
1 & 1 \\
1 & -1
\end{bmatrix}.
\end{aligned}
\]

The $X$-gate is a quantum analog of the NOT gate, while the Hadamard ($H$) gate places a qubit in superposition. Multi-qubit gates like the controlled-NOT (CNOT) enable entanglement by conditioning one qubit’s state on another
\[
\text{CNOT} = 
\begin{bmatrix}
1 & 0 & 0 & 0 \\
0 & 1 & 0 & 0 \\
0 & 0 & 0 & 1 \\
0 & 0 & 1 & 0
\end{bmatrix}.
\]

\subsection{Quantum Machine Learning}

Feature maps encode classical data $ \mathbf{x} \in \mathbb{R}^n $ into quantum states. The encoding is achieved through unitary transformations $ U(\mathbf{x}) $ that map data into a high-dimensional quantum Hilbert space. This process is formally represented as
\begin{equation}
\ket{\phi(\mathbf{x})} = U(\mathbf{x}) \ket{0}.
\end{equation}

For example, \emph{amplitude encoding}~\cite{Ventura_1999} embeds $ \mathbf{x} $ into the amplitudes of a quantum state, ensuring that $ \|\mathbf{x}\|^2 = 1 $. Other commonly used techniques in quantum machine learning include \emph{basis encoding}~\cite{Nielsen_Chuang_2010}, where each component of $ \mathbf{x} $ is mapped to computational basis states, and \emph{angle embedding}~\cite{Barenco_1995}, in which the components of $ \mathbf{x} $ are used as rotation angles in parameterized quantum gates (e.g., $R_y(x_i)$ or $R_z(x_i)$). Angle embedding is particularly hardware-friendly and suitable for near-term quantum devices \cite{Schuld_2021, Havl_ek_2019}.

An ansatz in quantum computing refers to a parameterized quantum circuit used to approximate solutions to optimization problems in QML. These circuits are constructed as layers of gates, with each gate having trainable parameters. The full circuit can be expressed as
\begin{equation}
U(\boldsymbol{\theta}) = \prod_{l=1}^L U_l(\boldsymbol{\theta}_l),
\end{equation}
where the parameters $ \boldsymbol{\theta} $ are optimized using classical algorithms to minimize a defined cost function. Common ansatz designs include hardware-efficient and problem-inspired ansätze \cite{Cerezo_2021}.

Quantum Machine Learning (QML) represents a convergent field that encompasses multiple paradigms at the intersection of quantum computing and machine learning. While one prominent approach leverages quantum computing principles to enhance classical machine learning capabilities, QML also includes the application of classical machine learning techniques to quantum data and systems. With the advent of quantum computing technologies, QML has gathered significant attention for its potential to address computationally intensive tasks and provide novel approaches to pattern recognition and data analysis \cite{Biamonte_2017, Carleo_2019}. At the core of QML implementations are Quantum Neural Networks (QNNs), which integrate quantum mechanical principles with neural network architectures through a hybrid quantum-classical framework. QNNs function by encoding classical information into quantum states and processing them through parameterized quantum circuits \cite{Benedetti_2019, Schuld_2020}. The hybrid nature of these systems is fundamental to their operation, as quantum circuits process the encoded data and generate quantum states that are subsequently measured to extract classical information. These measurement outcomes are then fed into a classical cost function, which is optimized using classical optimization algorithms that iteratively update the quantum circuit parameters until convergence is achieved. This quantum-classical feedback loop enables QNNs to leverage quantum phenomena such as superposition and entanglement while maintaining compatibility with classical machine learning frameworks, potentially achieving computational advantages over purely classical counterparts \cite{Liu_2021}.

A fundamental component in QML is the data re-uploading circuit architecture \cite{P_rez_Salinas_2020}. This approach alternates between encoding and processing unitaries, enabling the representation of non-linear functions through quantum operations while significantly increasing the expressivity of quantum models \cite{Schuld_2021}. Mathematically, a data re-uploading QNN with $L$ layers can be expressed as
\begin{equation}
\text{QNN}_{\boldsymbol{\theta}}(\mathbf{x}) = \prod_{l=1}^L U(\boldsymbol{\theta}_l)U(\mathbf{x}),
\end{equation}
where $U(\boldsymbol{\theta}_l)$ represents trainable unitary operations with parameters $\boldsymbol{\theta}_l$, and $U(\mathbf{x})$ encodes the input data $\mathbf{x}$. In practice, both $U(\boldsymbol{\theta}_l)$ and $U(\mathbf{x})$ are implemented as PQCs, typically consisting of layers of single-qubit rotations combined with entangling gates (such as CNOTs) to generate correlations between qubits.

The quantum embedding process maps classical data into quantum states through feature maps $\phi(\mathbf{x}) : \mathbb{R}^n \rightarrow \mathcal{H}$, where $\mathcal{H}$ represents the quantum Hilbert space. The encoding unitary $U(\mathbf{x})$ prepares a quantum state $\ket{\psi(\mathbf{x})}$ that corresponds to the classical input $\mathbf{x}$ under the chosen feature map, i.e.,
\begin{equation}
\ket{\psi(\mathbf{x})} = U(\mathbf{x})\ket{0}^{\otimes n} \approx \phi(\mathbf{x}),
\end{equation}
where the symbol $\approx$ reflects that the classical data may first be normalized or transformed before encoding it into the quantum state \cite{Havl_ek_2019}. This process enables access to high-dimensional Hilbert spaces for subsequent quantum processing. This quantum feature space offers several advantages:
\begin{itemize}
    \item \textbf{Enhanced expressivity}: QNNs can represent complex functions in quantum Hilbert spaces that might be inefficient to compute classically \cite{Abbas_2021}.
    \item \textbf{Quantum parallelism}: The ability to process multiple states simultaneously through quantum superposition \cite{Cerezo_2021}.
    \item \textbf{Efficient kernel computations}: Quantum circuits can compute kernel functions that might be classically intractable \cite{Huang_2021}.
\end{itemize}

\subsection{Advanced Quantum INR Models}
Mathematically, QIREN outputs a quantum-encoded function \( f(\mathbf{x}, \boldsymbol{\theta}) \) represented as
\begin{equation}
    f(\mathbf{x}, \boldsymbol{\theta}) = \langle 0 | U^\dagger(\mathbf{x}, \boldsymbol{\theta}) \mathcal{O} U(\mathbf{x}, \boldsymbol{\theta}) | 0 \rangle,
\end{equation}
where \( U(\mathbf{x}, \boldsymbol{\theta}) \) is a parameterized quantum unitary circuit that includes both input coordinates \( \mathbf{x} \in \mathbb{R}^d \) and trainable parameters \( \boldsymbol{\theta} \), and \( \mathcal{O} \) is a Hermitian observable. By leveraging Fourier properties inherent to quantum circuits, QIREN effectively represents the target function \( f(\mathbf{x}) \) as a sum of sinusoidal components
\begin{equation}
    f(\mathbf{x}) = \sum_{\omega \in \Omega} c_\omega e^{i \omega \cdot \mathbf{x}},
\end{equation}
where \( \Omega \) is the set of accessible frequencies determined by the eigenvalues of the encoding Hamiltonian \( H \), and \( c_\omega \) are coefficients controlled by the trainable gates and observable \( \mathcal{O} \).

The Quantum Fourier Gaussian Network (QFGN)~\cite{jin2025qfgnquantumapproachhighfidelity} augments the QIREN pipeline with a Fourier–Gaussian preprocessing layer. The input features \( \mathbf{x} \) are first projected onto a set of Fourier basis functions mixed with Gaussian weighting, represented mathematically as
\begin{equation}
    \mathbf{h}_1 = \Lambda \mathbf{B} \mathbf{x}_{\text{rep}} + \mathbf{b},
\end{equation}
where \( \mathbf{x}_{\text{rep}} \) is the repeated and concatenated input vector to create higher-dimensional representations, \( \mathbf{B} \) is the matrix of Fourier basis functions, \( \Lambda \) is a weight matrix, and \( \mathbf{b} \) is a trainable bias. The output \( \mathbf{h}_1 \) is then scaled by a Gaussian function to penalize the dominance of low-frequency components, represented as
\begin{equation}
    \varepsilon = \exp(-\gamma \mathbf{h}_1^2), \quad \mathbf{h}_2 = \varepsilon \mathbf{h}_1,
\end{equation}
where \( \gamma \) is a shaping parameter. The spectrally balanced features \( \mathbf{h}_2 \) are provided as input to the quantum encoding gates \( S(\mathbf{x}) = e^{-i H \mathbf{x}} \), where \( H \) (the Hamiltonian) contains tunable frequency components. The encoded quantum states are processed through a parameterized quantum circuit incorporating trainable quantum gates. The final output is derived via measurement of the observable \( \mathcal{O} \) to yield the representation \( f(\mathbf{x}) \).

These hybrid quantum–classical INR models leverage quantum Fourier features to compactly represent high-frequency structure~\cite{zhao2024quantumimplicitneuralrepresentations}, benefit from the expressive power of high-dimensional Hilbert-space embeddings~\cite{Schuld_2021}, and use parallel quantum operations that can accelerate nonlinear function approximation~\cite{schuld2020circuit}.

The applications of quantum implicit neural representations are broad and impactful across multiple domains. In image processing, these frameworks have been applied to adversarial generation of high-resolution images, outperforming classical Generative Adversarial Networks (GANs) in maintaining intricate details while requiring fewer parameters \cite{ma2025quantum}. Another application is in denoising diffusion models, where optimized quantum implicit models have been shown to improve noise reduction and image reconstruction tasks across datasets such as MNIST and CelebA \cite{ZHANG2025107875}. Similarly, in multimodal learning, the Quantum Multimodal Learning for Sentiment Classification (QMLSC) model demonstrates the utility of QINRs in integrating natural language and visual data for sentiment analysis, achieving higher classification accuracy than traditional multimodal neural architectures \cite{LI2025103049}. Quantum implicit representations have also been employed in compression tasks with the Quantum Implicit Neural Compression (quINR) framework \cite{fujihashi2024quantumimplicitneuralcompression}, which achieves superior rate-distortion trade-offs compared to classical codecs like JPEG2000. By leveraging the expressiveness of quantum neural networks, quINR encodes multimedia signals with smaller model sizes while retaining fine-grained details, enabling efficient storage and transmission of multimedia content \cite{fujihashi2024quantumimplicitneuralcompression}. Further applications include physics-informed modeling, video super-resolution, and other areas where high-dimensional data requires compact, continuous representations while preserving fidelity and minimizing resource overhead \cite{jin2025qfgnquantumapproachhighfidelity}.

\section{Detailed Q-NeRF Modules}
\label{sec:appendx_detailed_modules}

\begin{figure*}[tb]
    \centering
    \includegraphics[width=0.9\textwidth]{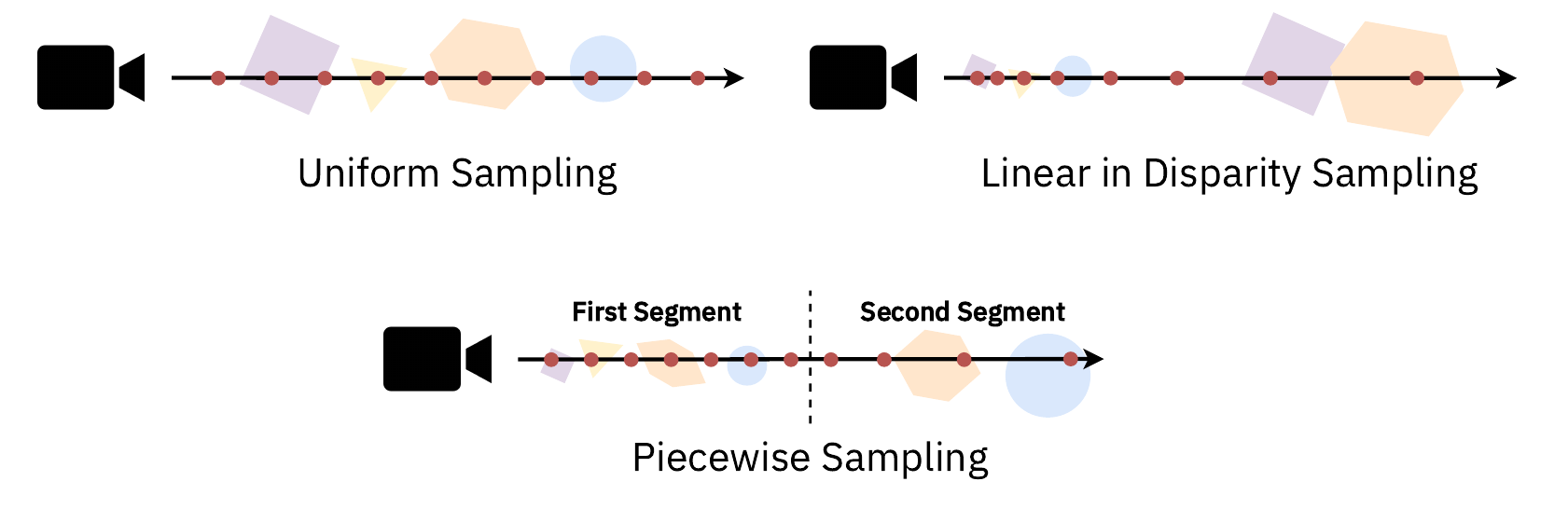} 
    \caption{
    Comparison of sampling strategies along a camera ray. The top row features uniform sampling (left) and linear in disparity sampling (right), and the bottom row illustrates the piecewise sampling strategy.
    }
    \label{fig:sampling_strategies}
\end{figure*}

\subsection{Pose Refinement}
\label{sec:pose_refinement}
Q-NeRF adopts the pose refinement procedure directly from the Nerfacto model to address inaccuracies in predicted camera poses, which are common when using consumer-grade devices such as smartphones. These inaccuracies, often present in poses derived from certain applications\footnote{For example, the Record3D iOS app is known to exhibit such issues.}, can result in misaligned 3D reconstructions and manifest as artifacts such as cloudiness, reduced sharpness, and loss of fine details. Leveraging the NeRF framework’s ability to backpropagate loss gradients, Nerfacto optimizes camera pose predictions to achieve accurate alignment of camera views, a strategy we retain in Q-NeRF.

At its core, pose refinement iteratively adjusts the camera's extrinsic parameters, specifically translation and rotation, within the SE(3) manifold. This optimization minimizes the reconstruction loss function, which accounts for discrepancies between synthesized views and corresponding ground truths. Defining the transformation matrix as
\begin{equation}
T = \left[ \begin{array}{cc} R & t \\ 0 & 1 \end{array} \right],
\end{equation}
where $R \in \mathbb{R}^{3\times3}$ is the rotation matrix and $t \in \mathbb{R}^{3}$ represents translation, the system leverages gradient descent to refine these parameters. The refinement process not only enhances per-frame alignment but also improves multi-view consistency, ensuring that spatial relationships between cameras are preserved. This advantage is particularly significant when reconstructing complex, high-variance scenes, as even minor pose errors can drastically impact rendering quality and scene coherence.

From a practical perspective, the integration of pose refinement in Nerfacto stands as a direct evolution from prior methods discussed in $\text{NeRF-}\text{-}$ \cite{wang2022nerfneuralradiancefields}. By employing pose optimization in the pipeline, the model offers superior resilience to noisy initial pose estimates, significantly reducing reliance on costly pre-calibration steps. Furthermore, accurate alignment translates into sharper renderings and reduced artifacts, as highlighted by the results from \cite{barron2022mipnerf360unboundedantialiased}.

\subsection{Piecewise Sampler}
\label{sec:picewise_sampler}

\begin{figure*}[tb]
    \centering
    \includegraphics[width=1\textwidth]{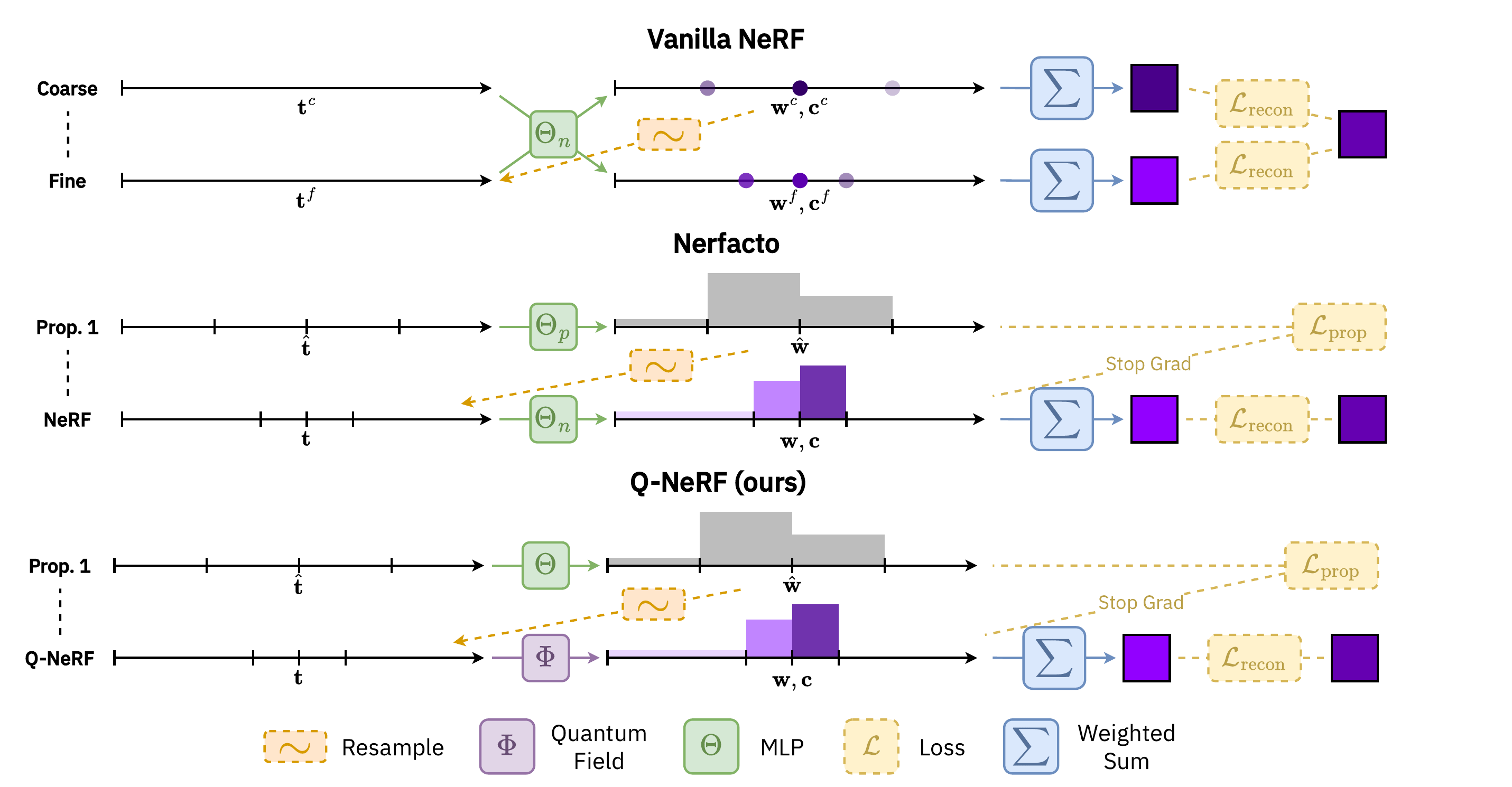} 
    \caption{Comparison of sampling strategies across Vanilla NeRF \cite{mildenhall2020nerfrepresentingscenesneural}, Nerfacto \cite{Tancik_2023}, and Q-NeRF (ours). The diagram illustrates the sequential processing pipelines, highlighting differences in proposal sampling, field representations, and supervision signals.}
    \label{fig:mipnerf_comparison}
\end{figure*}

The piecewise sampler is designed to initialize the sampling process efficiently by balancing sample density across varying object distances. As noted in \cite{Tancik_2023}, this method is particularly effective when modeling scenes that include both nearby and distant objects, ensuring sufficient detail capture throughout the spatial field. The primary role of the piecewise sampler is to generate an initial set of samples along rays cast through the scene. 

These samples are allocated in two distinct ways: half of the samples are distributed uniformly up to a fixed distance (e.g., 1 unit) from the camera, while the other half are allocated such that the step size increases progressively for distant objects. This hybrid approach combines the benefits of dense sampling for nearby details and adaptive sampling for far-away regions. For the latter half of the samples, the step size grows by scaling the sampling frustums iteratively. Mathematically, the step size, $\Delta_i$, for each sample point can be expressed as
\begin{equation}
\Delta_i = \Delta_0 \cdot s^i,
\end{equation}
where $\Delta_0$ is the base step size, $s > 1$ is the scaling factor, and $i$ indexes the sample along the ray. This geometric progression allows the sampler to allocate computational resources efficiently, reducing redundancy in sparse regions while maintaining fidelity in dense areas.

By employing this sampling strategy, the sampler satisfies two critical requirements. It captures high-frequency details in regions close to the camera, which is crucial for reconstructing fine-grained features and maintaining sharpness, and it also ensures that structures at farther distances are adequately represented without being oversampled, thus optimizing computational resources. This two-tiered sampling mechanism is closely related to principles outlined in Mip-NeRF \cite{barron2021mipnerfmultiscalerepresentationantialiasing}, which emphasize multi-scale sampling for efficient 3D scene modeling.

Figure~\ref{fig:sampling_strategies} illustrates various sampling strategies. The top left section depicts uniform sampling, where all sample points are placed at equal intervals along the ray, resulting in a uniform density of samples across the entire depth range. The top right section shows linear in disparity sampling, which adjusts the sample distribution to prioritize regions closer to the camera by decreasing the step size near the origin of the ray. The bottom part of the figure depicts the piecewise sampling strategy used in the Q-NeRF and Nerfacto models, included here for direct visual comparison. This visualization facilitates understanding of the optimization goals of various sampling methods employed in NeRF frameworks.

\subsection{Proposal Sampler}
\label{sec:proposal_sampler}
The proposal sampler is aimed at improving the precision and efficiency of scene reconstruction. It focuses on consolidating the sample locations along a ray into the regions of the scene that are most likely to contribute to the final rendered image, typically the first surface intersection. This targeted sampling significantly improves reconstruction quality by reducing the computational overhead involved in sampling irrelevant regions of the scene. To achieve this, it leverages a learned density function to guide the allocation of samples. This density function serves as a coarse approximation of the scene's true density distribution, providing a probabilistic map of where significant scene content is located.

\begin{figure*}[t!]
    \centering
    \includegraphics[width=\textwidth]{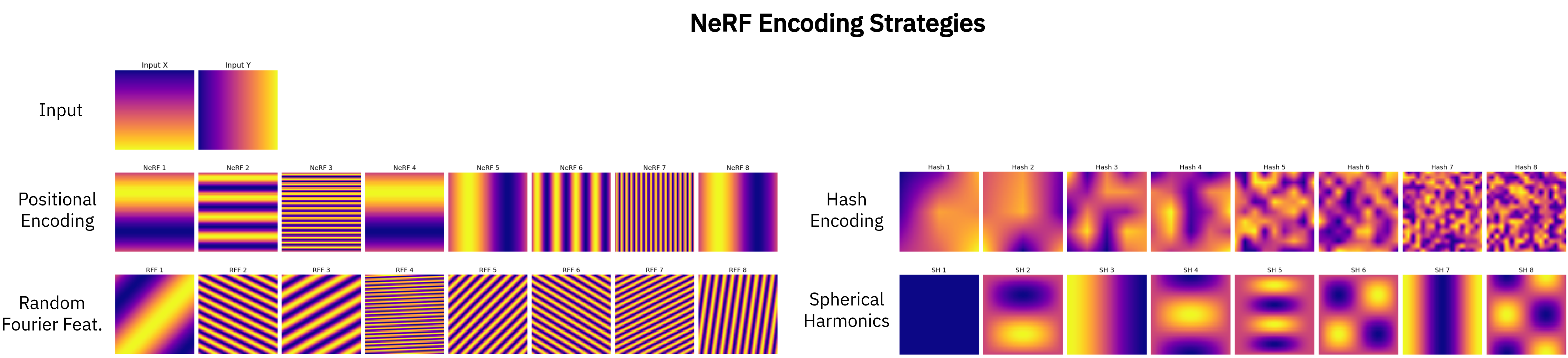}
    \caption{NeRF encoding strategies. The original spatial input coordinates are transformed through various encoding methods, including positional encoding, random Fourier features, hash encoding, and spherical harmonics. Each row visualizes the first eight channels produced by the respective encoding technique, illustrating differences in spatial frequency, locality, and smoothness.}
    \label{fig:nerf_encoding}
\end{figure*}

It operates hierarchically, chaining multiple density functions to refine the sampling regions iteratively. An initial coarse density function produces a set of candidate samples, which are subsequently refined by additional density functions to focus on the most relevant areas. Experimental results have shown that utilizing two density functions provides a significant improvement in sampling efficiency, while additional functions offer diminishing returns \cite{Tancik_2023}. Samples are allocated based on the probability density derived from the density functions. Specifically, the weights $w_i$ assigned to a sample $i$ are calculated as
\begin{equation}
w_i = T(t_i) \cdot \sigma_i \cdot (1 - e^{-\sigma_i \Delta_i}),
\end{equation}
where $T(t_i)$ is the accumulated transmittance along the ray up to sample $i$, $\sigma_i$ is the density at the sample location, and $\Delta_i$ is the sampling interval.

The density function, implemented as a lightweight fused multi-layer perceptron (MLP) with a hash encoding, efficiently estimates the density field of the scene, demonstrating the effectiveness of hash-based encodings in capturing scene details with minimal computation \cite{M_ller_2022}. The density function provides a smooth approximation of the scene, guiding the sampler toward relevant regions without requiring high-frequency details during initial passes. Formally, the density value $\sigma$ at a location $\mathbf{x}$ can be expressed as 
\begin{equation}
\sigma(\mathbf{x}) = f_\Theta(\mathbf{x}),
\end{equation}
where $f_\Theta$ is the fused MLP with hash-encoded input coordinates $\mathbf{x}$.

To ensure computational efficiency, the proposal sampler reduces the encoding dictionary size and the number of feature levels in the hash encoding. These adjustments have minimal impact on reconstruction quality since the density function does not need to capture high-frequency details. This design choice, combined with the hierarchical sampling strategy, allows for focusing computational resources on the regions that matter most for rendering. Its effectiveness is demonstrated in refining sample distributions and improving rendering quality across a variety of scenes \cite{barron2022mipnerf360unboundedantialiased}. By combining accurate density approximations with efficient sampling strategies, the proposal sampler achieves a balance between computational speed and visual fidelity.

As depicted in Figure~\ref{fig:mipnerf_comparison}, the proposal sampler implements a hierarchical sampling strategy that differs from the coarse-to-fine approach in other NeRF variants. It introduces an additional proposal stage that consolidates sampling regions using learned density functions, enabling efficient sampling while maintaining low computational overhead.

\subsection{Encoding Strategies}
\label{sec:nerfacto_field}
Figure~\ref{fig:nerf_encoding} provides a visual comparison of the classical encoding strategies used in NeRF pipelines. These encodings serve as the feature inputs to the density and color branches in the radiance field, and Q-NeRF interacts with them directly whenever a quantum module replaces the corresponding classical component.

\section{Extended Experimental Framework}
This section expands on the experimental design presented in the main paper, providing full details on hardware setup, dataset preprocessing, classical ablation studies, and the specific quantum configurations used in simulation.

\subsection{Dataset and Scene Configuration}
All experiments are conducted on the \textit{poster} scene from the Nerfstudio dataset collection.  
The full sequence contains 226 captured views, from which we obtain 204 training images and 22 held-out evaluation views following a 90/10 split. The original images (1080×1920 resolution) are downscaled to 36×64 pixels to ensure that quantum circuit simulation remains tractable. This resolution reduction is driven by the exponential time and memory requirements of state-vector simulation rather than by NeRF-specific constraints. Lower spatial resolution reduces the dimensionality of input encodings and the depth required for quantum circuits to capture spatial variation, making iterative optimization feasible within available compute resources. Camera poses and intrinsic parameters are taken directly from the Nerfstudio dataset without modification. No additional normalization, filtering, or temporal subsampling is applied, ensuring that all models, classical and hybrid, operate on identical and reproducible input data.

\subsection{Global Training Settings}\
All models are trained for a total of 30,000 iterations, following the optimization schedule described in the main paper.

The Adam optimizer is used with an initial learning rate of $0.01$ and an exponential decay schedule that follows a cosine-shaped ramp from a pre-warmup value of $1\!\times\!10^{-8}$ down to $10^{-4}$ by the end of training.  No warmup steps are applied.  To maintain consistent gradient statistics across all experiments, the proposal networks are updated every 5 iterations. Due to the memory footprint of state-vector quantum circuit simulation, the batch size is fixed at 128 rays per iteration for training and 64 rays for evaluation. Each of the two proposal stages uses 256 and 96 samples per ray, respectively, initialized via the piecewise sampler described in Appendix~\ref{sec:picewise_sampler}. Both proposal networks use five levels with a hidden dimension of 16, a $\log_2$ hash map size of 17, and maximum resolutions of 128 and 256. Proposal weight annealing is enabled using a slope of~10.0, with a warming period of 1,000 iterations.

\subsection{Hardware Configuration}
All experiments are conducted on the CVC's computing cluster equipped with dual-socket AMD EPYC 7352 24-core processors, yielding a total of 96 logical CPUs (48 physical cores with 2 threads per core). The cluster provides substantial computational resources, including $8\times$NVIDIA RTX 3090 GPUs (24GB VRAM each), with 32GB of RAM allocated per job. Each experiment in our study utilizes a single RTX 3090 GPU to accelerate the classical neural network components.

\subsection{Software Framework}
All quantum components are implemented using the PennyLane ecosystem, which provides differentiable quantum circuit simulation with seamless integration into PyTorch.

Quantum circuits are executed using PennyLane’s \texttt{default.qubit} simulator with full state-vector evolution, enabling backpropagation-based differentiation through the entire Q-NeRF pipeline.  This allows quantum parameters to be optimized jointly with classical network weights under a unified computational graph. Gradient evaluation is performed using analytic backpropagation rather than the parameter-shift rule, substantially reducing evaluation cost for small qubit counts. Several practical constraints arise from this setup. The exponential scaling of memory footprint requires careful control of circuit structure, particularly the number of qubits and data re-uploading layers used in each hybrid configuration. To ensure stable execution on GPU hardware, all simulations are performed in single precision with static computational graphs, and quantum circuits are compiled once at initialization to avoid runtime re-creation overhead.

\subsection{Classical Model Configuration Ablations}
\label{sec:classical_model_ablation}

\begin{table}[t]
\centering
\small
\renewcommand{\arraystretch}{1.2} 
\begin{tabular}{cccccccc}
\hline
\textbf{Config} & \textbf{L} & \textbf{m} & \textbf{M} & \textbf{H} & \textbf{F} & \textbf{MLP} & \textbf{Encoding} \\
\hline
C1 & 16 & 16 & 2048 & 19 & 2 & 2,502 & 16.7M \\
C2 & 16 & 16 & 2048 & 17 & 2 & 2,502 & 4.2M \\
C3 & 12 & 16 & 2048 & 17 & 2 & 1,990 & 3.1M \\
C4 & 12 & 16 & 2048 & 17 & 1 & 1,222 & 1.6M \\
C5 & 12 & 16 & 1024 & 15 & 1 & 1,222 & 393K \\
C6 & 8  & 16 & 1024 & 15 & 1 & 966   & 262K \\
C7 & 8  & 16 & 1024 & 13 & 1 & 966   & 66K \\
\hline
\end{tabular}
\vspace{0.7em}
\caption{Hash encoding configurations used in the ablation study. The table separates the number of parameters used by the MLP (\textit{MLP}) and by the encoding layer (\textit{Encoding}). L: number of levels; m: feature dimension per level; M: hash table size; H: number of resolution levels; F: fallback dimension.}
\label{tab:hash_encoding_configs}
\end{table}

To establish comprehensive baselines for Q-NeRF comparison, we conduct systematic ablation studies on classical Nerfacto architectures, exploring the impact of network depth, width, and encoding strategies on reconstruction quality. This analysis serves dual purposes: identifying optimal classical configurations within our resource constraints and providing reference performance metrics for evaluating quantum architectural innovations. Our ablation study systematically varies the multilayer perceptron (MLP) architecture used for density and color prediction while maintaining consistent training procedures and evaluation protocols. The exploration focuses on architectures feasible within our computational budget, ensuring that classical baselines represent achievable performance rather than idealized configurations beyond our experimental constraints.

\begin{figure*}[t!]
    \centering
    \begin{subfigure}[b]{0.48\textwidth}
        \centering
        \includegraphics[width=\textwidth]{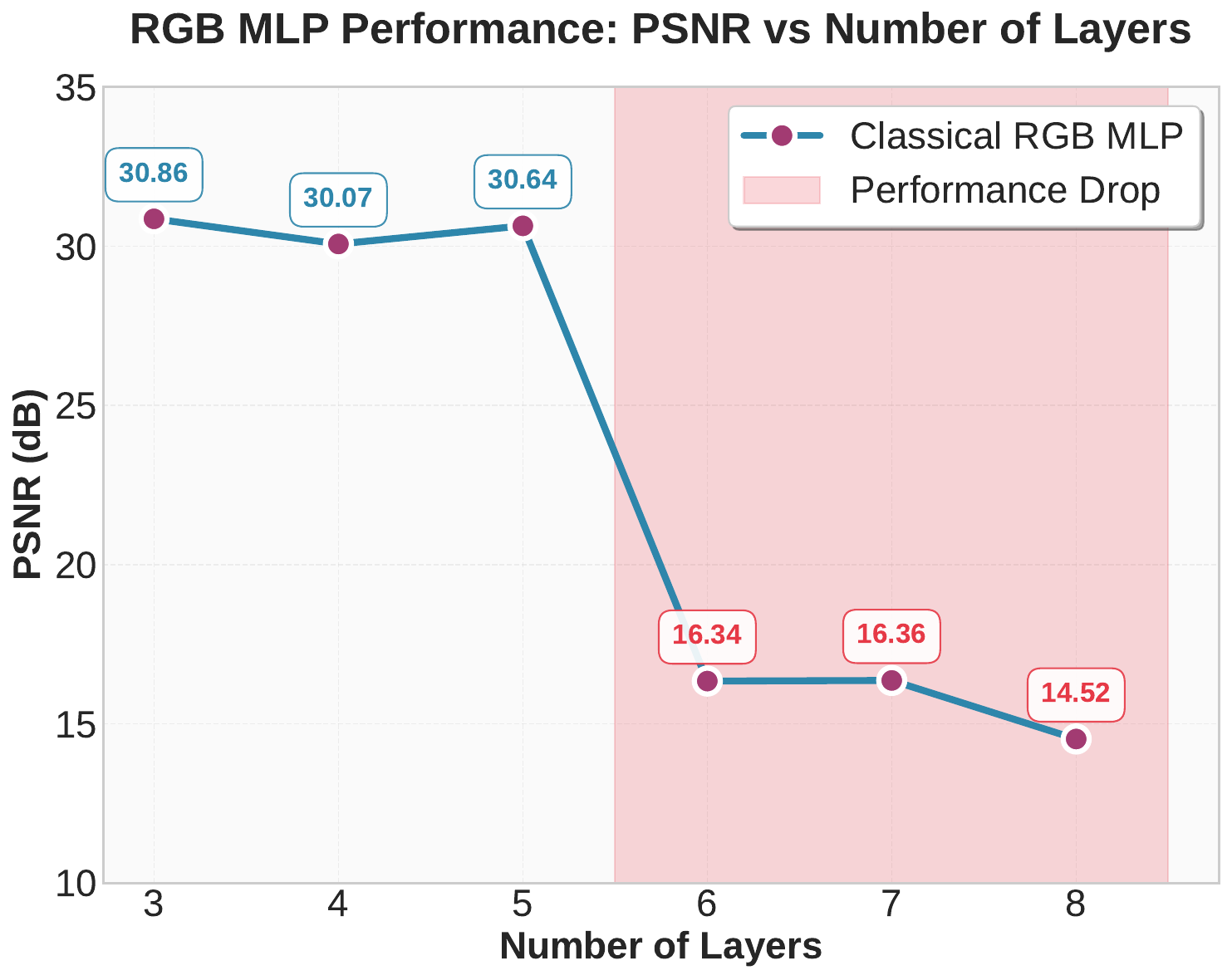}
        \caption{PSNR values for RGB MLPs with 3 to 8 layers (12 hidden units each). Each point represents the average PSNR across a validation set for a given network depth. The shaded region indicates a range of depths where performance changes noticeably.}
        \label{fig:classical_rgb_mlp_performance}
    \end{subfigure}
    \hfill
    \begin{subfigure}[b]{0.48\textwidth}
        \centering
        \includegraphics[width=\textwidth]{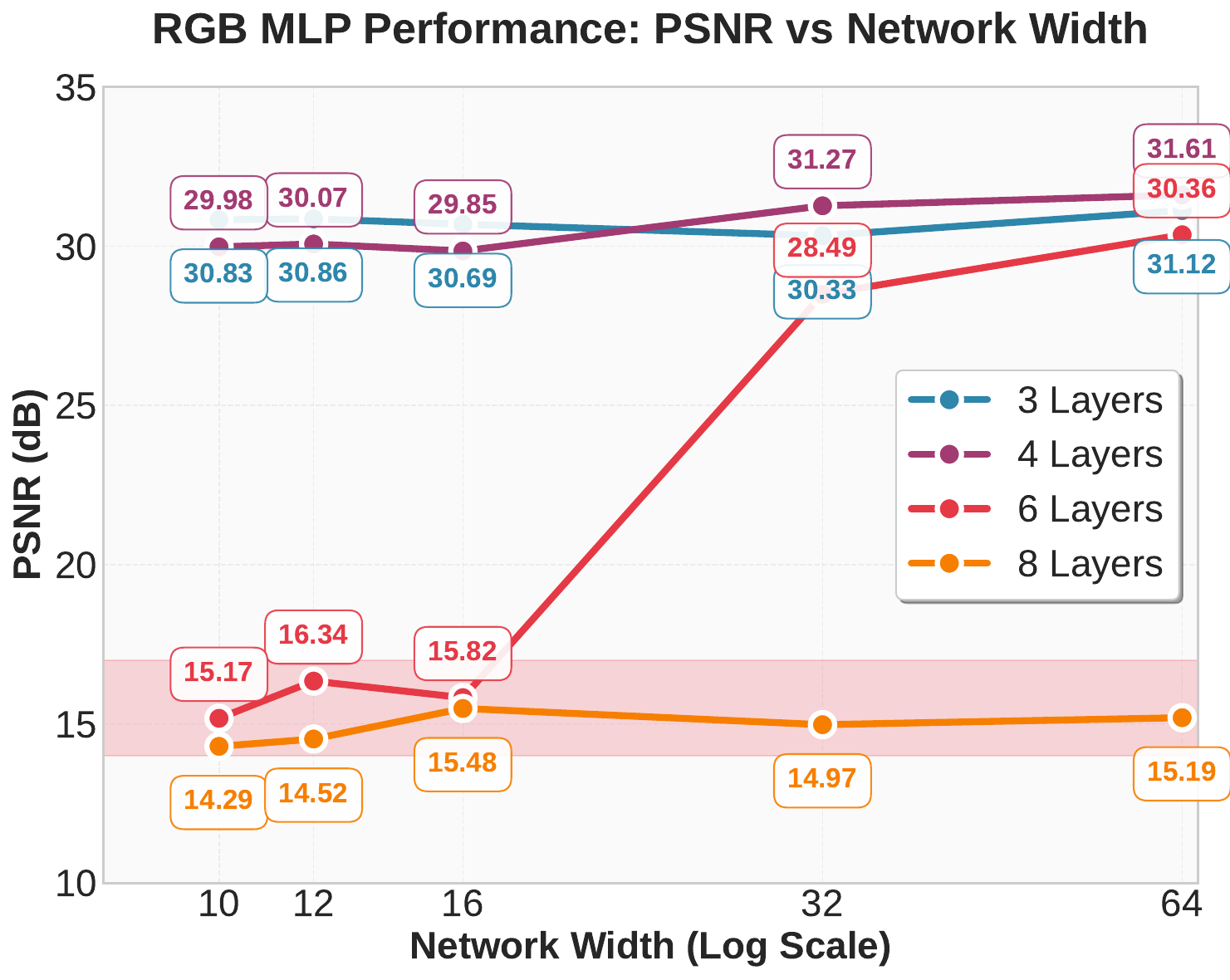}
        \caption{Performance comparison of RGB MLP networks with different layer configurations across varying network widths. PSNR performance in dB plotted against network width on a logarithmic scale from 10 to 64 hidden units.}
        \label{fig:rgb_psnr_vs_width}
    \end{subfigure}
    \caption{RGB MLP performance analysis: (a) Effect of network depth on performance with fixed width, and (b) Effect of network width on performance across different layer configurations for NeRF RGB color prediction. Results are based on a single run; since training involves random components such as parameter initialization, performance may vary across different seeds.}
    \label{fig:rgb_mlp_combined}
\end{figure*}

The systematic variation explores hidden layer counts ranging from 3 to 8 layers, with layer widths of 10, 12, 16, 32, or 64 neurons per layer. This parameter space spans architectures from compact models containing 363 parameters to larger networks with 26,563 parameters per MLP, offering a comprehensive overview of the classical performance landscape within our computational constraints. To isolate the effects of these architectural variations, certain components are held fixed. Specifically, the hyperparameters of the RGB color MLP are kept constant during density experiments, and the hyperparameters of the density MLP remain unchanged during RGB experiments, while the network parameters are still updated during training. This ensures a fair comparison by isolating the effect of modifying only one component at a time. Additionally, all other architectural elements remain fixed throughout, ensuring that no unintended modifications influence the results. The activation function is ReLU across all layers except the output, which uses a Sigmoid activation as specified in the standard Nerfacto configuration. This controlled design guarantees that observed performance differences arise solely from the varied architectural parameters under study.

The choice of spatial encoding strategy has a significant impact on both training efficiency and reconstruction quality in neural radiance fields. We conduct targeted ablation studies comparing hash encoding and positional encoding approaches within our classical baseline models, evaluating their effectiveness under the computational constraints imposed by our experimental setup.

For hash encoding experiments, we evaluate multiresolution hash tables with 13 to 19 resolution levels, hash table sizes ranging from 1024 to 2048 entries, and feature dimensions of 16 per level. The hash encoding configuration utilizes logarithmic spatial resolution scaling consistent with standard multiresolution hash grids, and employs a hash-based collision resolution strategy inherent to the encoding design. These parameters are selected to balance encoding expressivity with memory constraints, particularly considering the additional overhead introduced by quantum circuit simulation in subsequent experiments. Table \ref{tab:hash_encoding_configs} summarizes the complete configuration space, listing model identifiers, architectural parameters, encoding specifications, and total parameter counts for systematic comparison and reproducibility.

\begin{figure*}[t!]
    \centering
    \includegraphics[width=.75\textwidth]{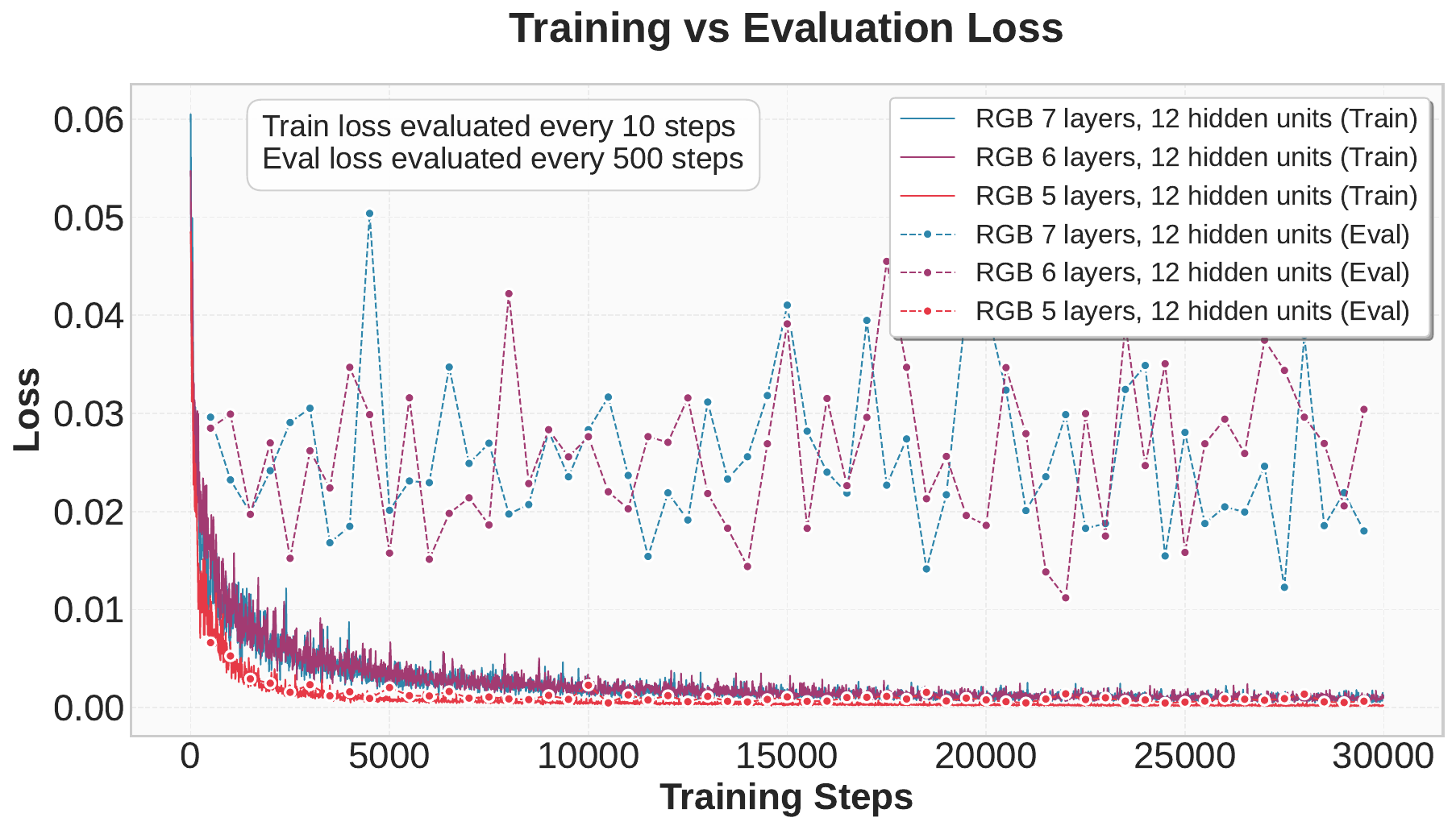}
    \caption{Training and evaluation loss for RGB prediction with 5, 6, and 7-layer MLPs (12 hidden units each). Solid lines represent the training loss evaluated every 10 steps; dashed lines represent the evaluation loss evaluated every 500 steps. Each color corresponds to a different network depth as indicated in the legend.}
    \label{fig:train_eval_loss}
\end{figure*}

Positional encoding ablations explore sinusoidal encodings with frequency ranges from $2^{0}$ to $2^{8}$, applying 60 encoding dimensions to the 3D spatial coordinates. The positional encoding depth is fixed at 10 frequency bands, reflecting a consistent encoding size across all experiments to isolate the effects of network capacity. Layer width variations in positional encoding experiments range from 12 to 64 neurons, compensating for the fixed-size encoding input through adjustments in the network’s representational capacity.

\subsection{Classical Baseline Results}
Before analyzing the hybrid quantum–classical architecture, it is important to first establish a baseline understanding of the classical RGB MLP performance. Since the hybrid classical-quantum layers are designed to replace or augment the classical RGB head, evaluating how depth and width affect a purely classical MLP provides critical context. This exploration allows us to disentangle limitations that arise from the inherent design of the RGB head itself from those introduced by the quantum components, ensuring that observed performance differences in the hybrid model are not simply artifacts of under- or over-parameterized classical baselines.

Figure~\ref{fig:classical_rgb_mlp_performance} shows that PSNR remains consistent for RGB MLPs with 3 to 5 layers, but drops sharply when the depth exceeds 5 layers. This degradation in performance for deeper networks suggests that increasing the number of layers in the classical RGB head does not improve expressiveness and may instead introduce optimization challenges. A plausible explanation is the limited capacity of the shallow SH-encoded directional input, which becomes increasingly difficult to reconcile with deeper RGB representations. This misalignment may hinder effective gradient flow or amplify sensitivity to high-frequency noise, pointing to the need for architectural adjustments that better integrate directional cues at greater depths.

The relationship between network width and performance, shown in Figure~\ref{fig:rgb_psnr_vs_width}, reinforces the findings from the depth analysis while revealing additional architectural insights. Consistent with the previous observations, 3-layer and 4-layer networks maintain high performance across all tested widths, with 4-layer networks achieving the peak PSNR of 31.6 dB at 64 hidden units. The 6-layer configuration exhibits a dramatic performance recovery as width increases, jumping from 16.3 dB at width 12 to 28.5 dB at width 32, suggesting that the optimization challenges identified in deeper networks can be partially mitigated through increased representational capacity. However, this improvement comes at a significant computational cost and still falls short of the consistent performance delivered by shallower architectures. The 8-layer networks remain persistently poor performers regardless of width, indicating that beyond a certain depth threshold, the fundamental architectural limitations cannot be overcome through increased width alone. This pattern supports the hypothesis that the mismatch between shallow SH-encoded directional inputs and deep RGB representations creates optimization bottlenecks that are more effectively addressed through architectural redesign rather than simple capacity scaling.

\begin{figure}[t!]
    \centering
    \includegraphics[width=0.47\textwidth]{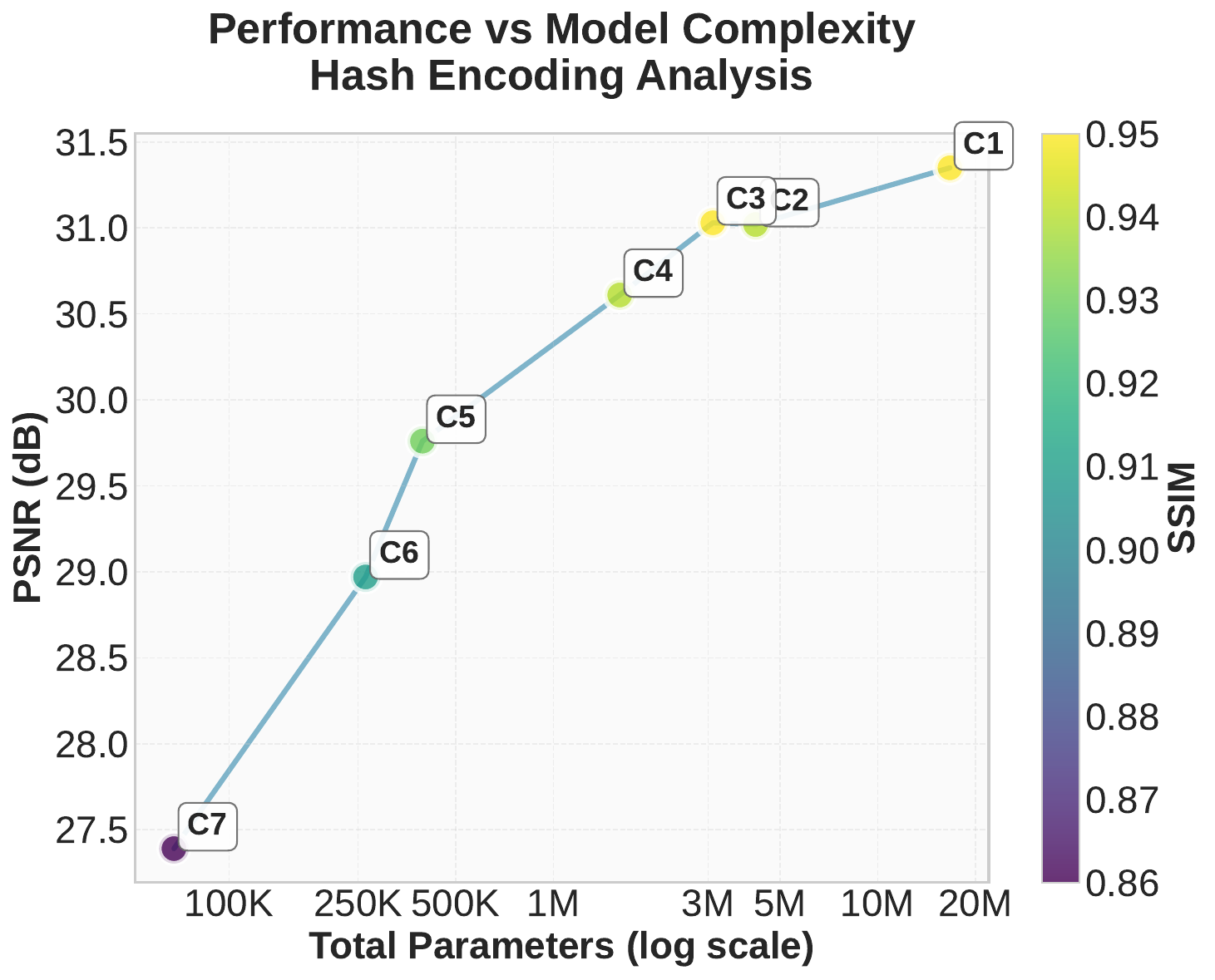}
    \caption{Effect of network width on performance using multiresolution hash encoding in the density MLP. 
    Each point corresponds to a specific hash encoding configuration (C1–C7), with PSNR plotted against total parameter count.
    Marker color indicates SSIM.}
    \label{fig:hash_encoding_analysis}
\end{figure}

\begin{figure*}[t!]
    \centering
    \includegraphics[width=1\textwidth]{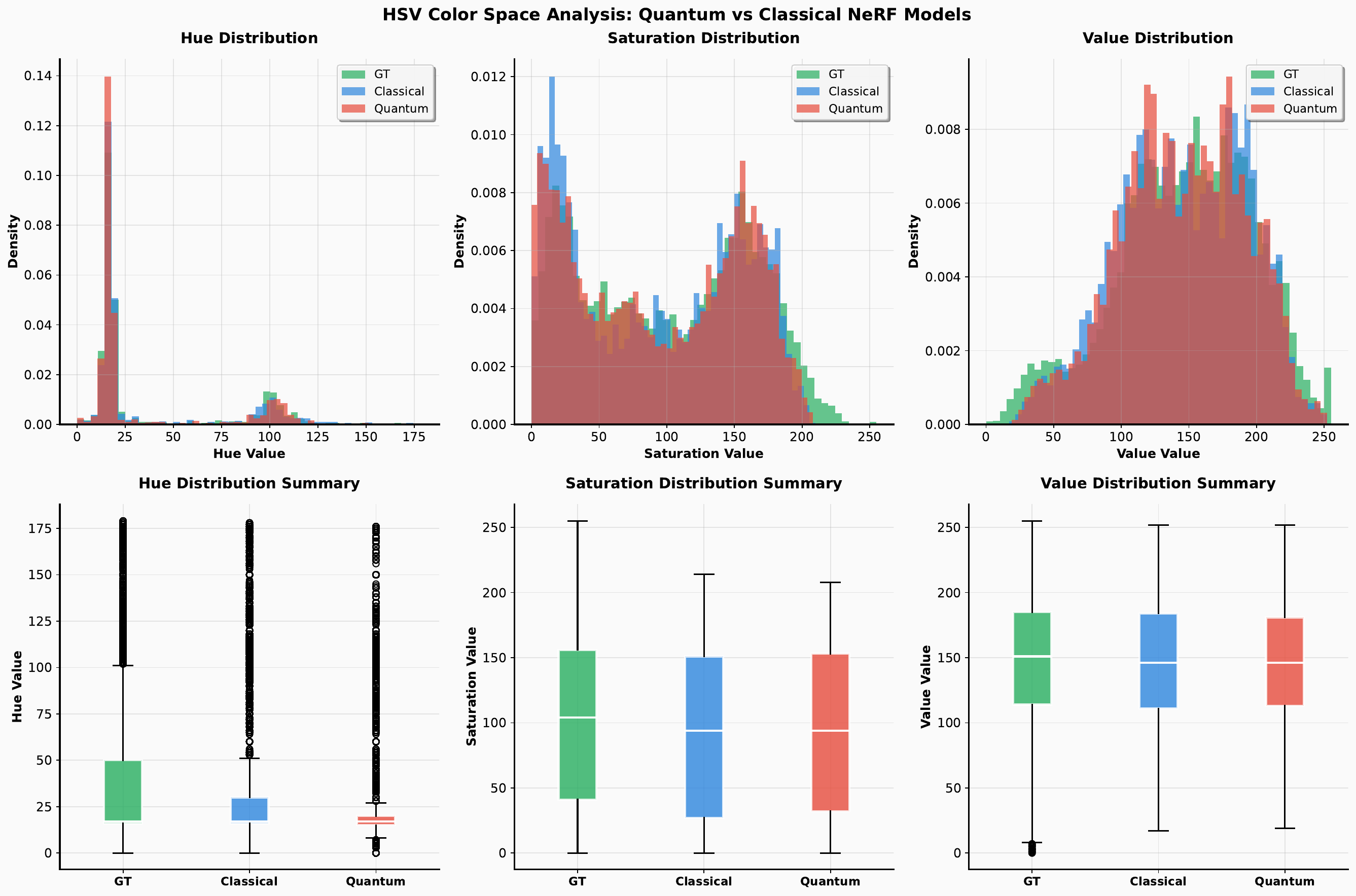}
    \caption{HSV color space analysis comparing Ground Truth (GT), a Classical NeRF baseline (Nerfacto with positional encoding; Density Net: Classical MLP with 1,200 parameters, Color Net: Classical MLP with 531 parameters, 4 density + 3 RGB layers, width 12), and a Q-NeRF variant (Density Net: QIREN with 1,166 parameters + positional encoding, Color Net: Classical MLP with 1,859 parameters, $2L\times2S$ density layers + 3 RGB layers, width 8 + 32). The top row shows distributions of Hue, Saturation, and Value, while the bottom row presents corresponding boxplot summaries.}
    \label{fig:hsv_analysis}
\end{figure*}

A similar pattern emerges in Figure~\ref{fig:train_eval_loss}, where deeper MLPs demonstrate lower training loss but significantly higher and more unstable evaluation loss. This gap between training and evaluation behavior indicates a heightened tendency to overfit as depth increases. In contrast, shallower configurations, particularly the 5-layer model, maintain smooth and stable validation curves. Taken together, these results suggest that deeper classical RGB heads can compromise generalization and stability, underscoring the importance of reevaluating how directional encoding is handled in deeper architectures.

Figure~\ref{fig:hash_encoding_analysis} presents the performance of density MLPs trained with multiresolution hash encoding.  
Across the explored configurations (C1–C7), hash-based models consistently achieve strong reconstruction fidelity, with peak performance at 31.35~dB PSNR and 0.95~SSIM for the largest model (C1).  
Several mid-sized variants (e.g., C3 and C4) exceed 31~dB PSNR while using substantially fewer parameters, confirming the effectiveness of spatial hashing in capturing high-frequency geometric details even under low-resolution and small-batch training constraints.

The most efficient configuration, C7, attains competitive performance with fewer than 100{,}000 total parameters, demonstrating the adaptability of hash features to compact density networks.  
The exact architectural specifications corresponding to C1–C7 are provided in Table~\ref{tab:hash_encoding_configs}.

\section{HSV Output Analysis}
\label{sec:appendix_hsv_output}

In this experiment, we evaluate the color reproduction fidelity of two NeRF variants against the ground truth (GT) by analyzing their outputs in the HSV (Hue, Saturation, Value) color space. The motivation for this analysis is to move beyond pixel-wise error metrics and instead assess the perceptual quality of the generated images through a distributional comparison of their color components. To this end, we compare a classical NeRF baseline with positional encoding and a Q-NeRF variant, in which the density field is modeled using a hybrid classical-quantum representation while color is rendered via a classical MLP.  

The classical NeRF configuration consists of a density network implemented as a classical MLP with 1,200 parameters and a color network with 531 parameters, organized into four density layers and three RGB layers with width 12. The Q-NeRF configuration, in contrast, employs a density network implemented using QIREN with 1,166 parameters and positional encoding, and a color network modeled by a classical MLP with 1,859 parameters, structured as $2L\times2S$ density layers and three RGB layers with widths of 8 and 32, respectively. Rendered outputs from both models were converted into HSV color space, and histograms of hue, saturation, and value were computed over all pixels. To complement the distributions, we also provide boxplot summaries, as shown in Fig.~\ref{fig:hsv_analysis}.  

The hue distributions reveal a sharp peak in the ground truth below a value of 25, corresponding to a strong concentration of low-hue colors. The classical NeRF reproduces this peak but exhibits a wider spread around it, indicating less stability in hue representation. The Q-NeRF variant also captures the peak and aligns well with the ground truth in terms of central tendency; however, its boxplot reveals a larger number of outliers across the hue range. This suggests that while Q-NeRF preserves the dominant hue structure more accurately, it introduces more spurious deviations at the distribution tails compared to the classical model.

Saturation analysis shows that the ground truth spans a broad range, with a concentration in the mid-to-high saturation interval (100–200), and in the low interval (0-50). Both classical and Q-NeRF models successfully approximate this bimodal distribution pattern. Both models closely replicate the ground truth's saturation characteristics, showing similar density profiles across the low saturation peak (0-50) and the broader mid-to-high saturation region (100-200). The box plots further confirm this similarity, with both models displaying comparable median values and quartile ranges to the ground truth. This indicates that both approaches effectively preserve the color saturation properties of the original scenes, with no significant advantage of one model over the other in terms of saturation reproduction.

For the value component, the ground truth distribution is centered between 100 and 200, forming a bell-shaped profile. Both the classical NeRF and Q-NeRF closely approximate this shape and central tendency. The distributions from both models are remarkably similar to each other and to the ground truth, with comparable spread and centering. The box plots confirm this similarity, showing that both models achieve nearly identical median values and quartile ranges that closely match the ground truth. Both approaches demonstrate equivalent performance in preserving the brightness characteristics of the original scenes. 

We performed Kolmogorov–Smirnov (KS) tests to quantify distributional differences between ground truth (GT), classical-rendered, and quantum-rendered images in the HSV color space. For the \textit{hue} channel, both GT vs. classical (KS = 0.0342, $p < 10^{-5}$) and GT vs. quantum (KS = 0.0805, $p < 10^{-6}$) comparisons showed highly significant differences, with classical vs. quantum also highly significant (KS = 0.0649, $p < 10^{-6}$). In the \textit{saturation} channel, GT vs. classical (KS = 0.0749, $p < 10^{-6}$) and GT vs. quantum (KS = 0.0490, $p < 10^{-6}$) were both highly significant, as was classical vs. quantum (KS = 0.0338, $p < 10^{-5}$). For the \textit{value} channel, all comparisons were again highly significant: GT vs. classical (KS = 0.0367, $p < 10^{-5}$), GT vs. quantum (KS = 0.0384, $p < 10^{-6}$), and classical vs. quantum (KS = 0.0285, $p < 10^{-3}$). These results indicate that quantum rendering induces statistically measurable deviations from both classical outputs and GT images across all HSV channels.

Taken together, the HSV analysis highlights that both Q-NeRF and classical NeRF achieve comparable alignment with the ground truth across all color dimensions. Both models demonstrate similar performance in reproducing hue, saturation, and brightness characteristics, with no clear advantage of one approach over the other. The distributions show that both methods effectively preserve the chromatic and luminance properties of the original scenes, suggesting that both quantum-enhanced and classical density encoding approaches are equally capable of maintaining perceptual color accuracy in NeRF rendering.

\end{document}